\documentclass[fleqn,usenatbib]{mnras}

\usepackage{newtxtext,newtxmath}
\usepackage{soul}
\usepackage[T1]{fontenc}
\DeclareRobustCommand{\VAN}[3]{#2}
\let\VANthebibliography\thebibliography
\def\thebibliography{\DeclareRobustCommand{\VAN}[3]{##3}\VANthebibliography}

\usepackage[dvipsnames]{xcolor}

\usepackage{graphicx}
\usepackage{amsmath}
\usepackage{amssymb}
\usepackage{changes}
\usepackage{hyperref}

\usepackage{adjustbox}
\usepackage{anyfontsize}
\usepackage{longtable}
\usepackage{afterpage}

\let\oldAA\AA
\renewcommand{\AA}{\text{\normalfont\oldAA}}

\title[Spatial correlation between Quaia AGN and GWs]
{Constraining the AGN formation channel for detected black
hole binary mergers up to z=1.5 with the Quaia catalogue}

\author[N. Veronesi et al.]{
Niccolò Veronesi,$^{1}$\thanks{E-mail: veronesi@strw.leidenuniv.nl}
Sjoert van Velzen,$^{1}$
Elena Maria Rossi,$^{1}$
Kate Storey-Fisher,$^{2}$
\\
$^{1}$ Leiden Observatory, Leiden University, PO Box 9513, 2300 RA Leiden,
The Netherlands\\
$^{2}$ Donostia International Physics Center, Manuel Lardizabal Ibilbidea,
4, 20018 Donostia, Gipuzkoa, Spain}
\date{Accepted 2024 November 11. Received 2024 October 29; in original form 2024 July 31}

\pubyear{2023}

\begin{document}
\label{firstpage}
\pagerange{\pageref{firstpage}--\pageref{lastpage}}
\maketitle

\begin{abstract}
Statistical analyses based on the spatial correlation between the sky maps of
Gravitational Wave (GW) events and the positions of potential host environments
are a powerful tool to infer the origin of the black hole binary mergers that have
been detected by the LIGO, Virgo, and KAGRA instruments. In this paper, we
tighten our previous constraints on the fraction of detected GW events that may
have originated from Active Galactic Nuclei (AGN). We consider 159 mergers detected
not later than June 1st, 2024, and the all-sky quasar catalogue Quaia. We
increase by a factor of 5.3 and 114 the number of considered GW sources and AGN
respectively, also extending our analysis from redshift $0.3$ to $1.5$. This is
possible thanks to the uniformity of the AGN catalogue and its high level of
completeness, which we estimate as a function of redshift and luminosity. We
find at a 95 per cent credibility level that un-obscured AGN with a bolometric
luminosity higher than $10^{44.5}{\rm erg\ s}^{-1}$ ($10^{45}{\rm erg\ s}^{-1}$)
do not contribute to more than the 21 (11) per cent of the detected GW events.
\end{abstract}



\section{Introduction}
\label{sec:intro}

With the only exception of the Gravitational Wave (GW) event GW170817
\citep{Abbott17}, a confident one-to-one association between the mergers
of compact objects detected by the LIGO-Virgo-KAGRA (LVK) collaboration
\citep{Acernese15,LIGO15,Akutsu21} and their host galaxy is currently out
of reach.

This is primarily caused by the typical size of the uncertainties that are
associated to the sky position and to the luminosity distance of each GW
event. The difficulty of associating each merger to its host environment
hampers our ability to pin down the physical origin of the binary systems
of which the coalescences have been directly measured. In one class of the
possible formation scenarios that have been proposed
\citep[see][for a review]{Mapelli21} binaries of compact objects like
binary black holes, binary neutron stars, and neutron star-black hole binaries
are efficiently assembled and driven to merger in dense environments like
globular clusters \citep{rodriguez16} or nuclear stellar clusters
\citep{Chattopadhyay23}. In such hosts, binaries can have their semi-major axes
shrunk by interactions with single objects. This binary hardening from
single-binary encounters is important for the formation of systems that are
able to merge within a Hubble time due to loss of orbital energy, emitted in
the form of GWs. The least massive object of the three is expected to be ejected
from the location of the single-binary interaction \citep{Hills80,Ziosi14},
meaning that the hardened remnant binary will be composed by the two heaviest
elements that partake in the encounter.

Another reason why dynamically dense environments might be the hosts of the
most massive mergers detected by the LVK collaboration is the fact that their
escape velocities can be large enough to retain the recoil-kicked remnant
of a coalescence, turning it into a potential component of a subsequent GW event.
This ``hierarchical merger” scenario might be a physical interpretation
for the existence of stellar-mass Black Holes (sBHs) in the pair-instability
mass gap \citep{Gerosa17,Gerosa19,yang19,barrera22}. This discontinuity is
predicted between $\approx50{\rm M}_\odot$ and $\approx120{\rm M}_\odot$
in the sBH mass spectrum as a consequence of the complete disruption of the
core during the supernova event at the end of the life cycle of stars with
very high masses, that are therefore expected to leave no remnant
\citep{Heger02,Belczynski16}. However, the astrophysical mass distribution
of sBHs predicted by the LVK collaboration shows evidence for the presence
of objects in the pair-instability mass gap \citep{Abbott23}. This
challenges the hypothesis that all the merging systems detected by the LVK
collaboration have originated from an isolated stellar binary
\citep[however, see also][]{Belczynski20}.

Accretion discs of AGN are a unique type of potential host environment for the
assembly and the merger of binaries of compact objects. The reason is that in
this so-called ``AGN formation channel” the binaries and their components
are not expected to interact only with other stellar objects, but also with
the gas the accretion disc consists of. Thanks to this interaction, the disc
might for example capture compact objects that have orbits moderately inclined
with respect to its plane \citep{Ostriker83,Fabj20,Nasim23}. These disc captures
have the effect of increasing the number density of compact objects in the disc,
where the binary formation can be gas-assisted \citep{Bartos17a,Tagawa20,DeLaurentiis23,Rowan23}.
Another process which, as far as the formation of merging binaries of compact
objects is concerned, is typical of the AGN scenario is migration: the radial
motion of compact objects orbiting around the central supermassive black hole.
Inward migration takes place in AGN discs where the net torque exerted by the
gas onto the orbiting compact object is negative \citep{Paardekooper10,Bellovary16},
and can increase the number density of sBHs and neutron stars in the inner
part of the accretion disc, facilitating the formation of binaries.

One possible way to put constraints on the fractional contribution of the AGN
channel to the total merger rate of compact object binaries is through the
investigation of the spatial correlation between the sky maps of the events
detected by the LVK collaboration and the positions of observed AGN. This approach
has been first suggested in \citet{Bartos17}, where it was estimated that 300
GW detections are needed to statistically prove, with a significance of $3\sigma$,
the GW-AGN connection if half of the mergers happened in an AGN. In
\citet{corley19} it was later found that the number of required detections
decreases by a factor $\approx3$ if it is assumed that the inclination of the
binary angular momentum with respect to the line of sight is known with an
uncertainty of $5^\circ$. In the simulated GW detections used in
\citet{Bartos17} and \citet{corley19} the interferometers of LIGO and
Virgo have been assumed to be working at design sensitivity. This level of
instrumental sensitivity is a factor $\gtrsim2$ better with respect to the one reached
by the interferometers during the first three observing runs \citep{Cahillane22}.

In \citet{veronesi22} the analysis presented in \citet{Bartos17} was repeated
assuming a realistic distribution of the sizes of the 90 per cent Credibility Level (CL) localisation
volumes (V90) of the mock GW detections. This distribution has been created by
using the sensitivity curves that characterised the LIGO and Virgo interferometers
during their third observing run (O3). It was found that the amount of data
collected during O3 would be enough to prove with a $3\sigma$ significance that
the AGN channel contributes to half of the total merger rate only if the GW
events originate form a rare host population with a number density lower than
$10^{-7}{\rm Mpc}^{-3}$. Integrating the the AGN luminosity function of
\citet{Hopkins07}, we find that this density corresponds to objects in the
local Universe with a bolometric luminosity greater than approximately
$10^{46}{\rm erg\ s}^{-1}$.

By building on this previous works, the first observational constraint on the
efficiency of the AGN channel was put using the spatial-correlation approach
in \citet{Veronesi23}, where the completeness of the quasar catalogues used
during the cross-match with GW sky maps and the exact position of the
potential host environments have been taken into account for the first time.
By using catalogues of AGN with a spectroscopic estimate of redshift of $z\leq0.3$
obtained from Milliquasv7.7b \citep{flesch21} and the sky maps of the 30 GW
events detected in the same redshift range during the first three observing runs
of the LVK collaboration, it was found that the fraction of the detected mergers
that have originated in an AGN more luminous than $10^{45.5}{\rm erg\ s}^{-1}$
($10^{46}{\rm erg\ s}^{-1}$) is not expected to be higher than 0.49 (0.17) at a CL of 95 per cent.
A similar statistical investigation has been conducted also in \citet{Veronesi24}.
In this case the sky maps of the GW events detected during O3 have been cross-matched
spatially and temporally with the 20 unusual AGN flaring activities detected by
the Zwicky Transient Facility \citep{Bellm19,Graham19} that have been identified
in \citet{Graham23} as potential transient electromagnetic counterparts. We find no
evidence for a correlation the GW events and the unusual flares. A similar result
has been found in \citet{Palmese21} analysing the spatial correlation between the
event GW190521 and an unusual flaring activity of the AGN J124942.3+344929.

In this work we present new observational constraints on the fractional contribution
of the AGN channel to the total merger rate, $f_{\rm AGN}$. We use a
spatial-correlation-based method similar to the one presented in \citet{Veronesi23},
and apply it to a larger dataset. In particular we use the sky maps of the
GW events detected by the interferometers of the LVK collaboration up to June 1st,
2024, therefore including data coming from the fourth observing run (O4). This
brings the total number of used sky maps to 159, which is more than five times
the amount of GW events used in \citet{Veronesi23}. These are cross-matched with
the AGN of the all-sky catalogue Quaia \citep{StoreyFisher24}, which is derived
from the catalogue of extra-galactic quasar candidates published by the Gaia mission
\citep{Gaia23}. The remarkable uniformity and completeness up to $z=1.5$ of Quaia
make it a relevant tool for all-sky spatial correlation analyses like the one
presented in this work.

In Section \ref{sec:dataset} we present the main characteristics of our
dataset, which includes the different GW detections and the Quaia
catalogue. In Section \ref{sec:method} we then describe the likelihood
function that we maximise in order to obtain our new constraints on
$f_{\rm AGN}$, which are presented in Section \ref{sec:res}. Finally, in
Section \ref{sec:concl}, we draw the main conclusions regarding the AGN
channel that can be inferred from the results of our analysis, and discuss
about future developments of statistical spatial-correlation-based approaches.
We adopt the cosmological parameters of the Cosmic Microwave Background
observations by Planck \citep{Planck16}:
$H_0=(67.8\pm0.9)\ {\rm km}\ {\rm s}^{-1}{\rm Mpc}^{-1}$,
$\Omega_{\rm m}=0.308\pm0.012$, and $n_{\rm s}=0.968\pm0.006$.


\section{Datasets}
\label{sec:dataset}
In this section, we first present the main characteristics of the GW events
the sky maps of which are used in our analysis. We then describe the main
properties of the Quaia catalogue and how we estimate its completeness
as a function of redshift for different sub-samples of it, characterised
by different thresholds of bolometric luminosity, $L_{\rm bol}$.


\subsection{GW events}
\label{sec:GWs}
In this work we use the 159 sky maps of mergers of compact object binaries
detected by the interferometers of the LVK collaboration up to June 1st,
2024. In particular, all the GW events from the first three observing runs
are included, with the exclusion of the binary neutron star merger
GW170817 and of the two binary black hole mergers GW200308\_173609 and
GW200322\_091133. While the former has not been considered our analysis
since its host galaxy has been identified thanks to the direct detection
of its electro-magnetic counterpart \citep{Abbott17}
\footnote{We decide to use in our analysis only detected GW events
which lack a confident host galaxy identification. However, the
non-inclusion of GW170817 is not expected to have any significant
impact on our final results regarding the whole population of detected
mergers.}, the latter events have not been taken into account because
they are poorly localised, and the corresponding value of V90 cannot be
estimated from the currently available posterior samples. Moreover, we
use all the events from the first half of O4 (O4a) as well as all the
events from its second half (O4b) that have been detected not later than
June 1st, 2024. For this currently ongoing observing run, we select the
detections that have a probability of being of terrestrial origin smaller
than 1 per cent.

The sky maps of the events detected during the first three observing
runs are downloaded from the Gravitational Wave Open Science Center
\citep{Abbott23}. We use the posterior samples obtained
using the {\scshape IMRPhenomXPHM} waveform model \citep{Pratten21}
for all these events but GW190425\_081805, GW191219\_163120,
GW200105\_162426, and GW200115\_042309. For these four events
we use the {\scshape Mixed} posterior samples.

The sky maps of the 71 events of O4 are downloaded from the Gravitational-Wave
Candidate Event Database \footnote{\url{https://gracedb.ligo.org/}}, operated
by the LIGO Scientific Collaboration. For each event, we use the most recent
version of its sky map, which is either obtained from the Bilby localisation
algorithm \citep{Ashton19} or from the Bayestar one \citep{Singer16}. We here
specify as a caveat that these maps consist of early estimates, and
may differ from the ones that will be published in the future Gravitational-wave
Transient Catalogue (GWTC) containing all the properties of the events
detected during O4.

\begin{figure}
    \centering
    \includegraphics[trim= 0 0 0 0 ,clip,width=1\columnwidth]{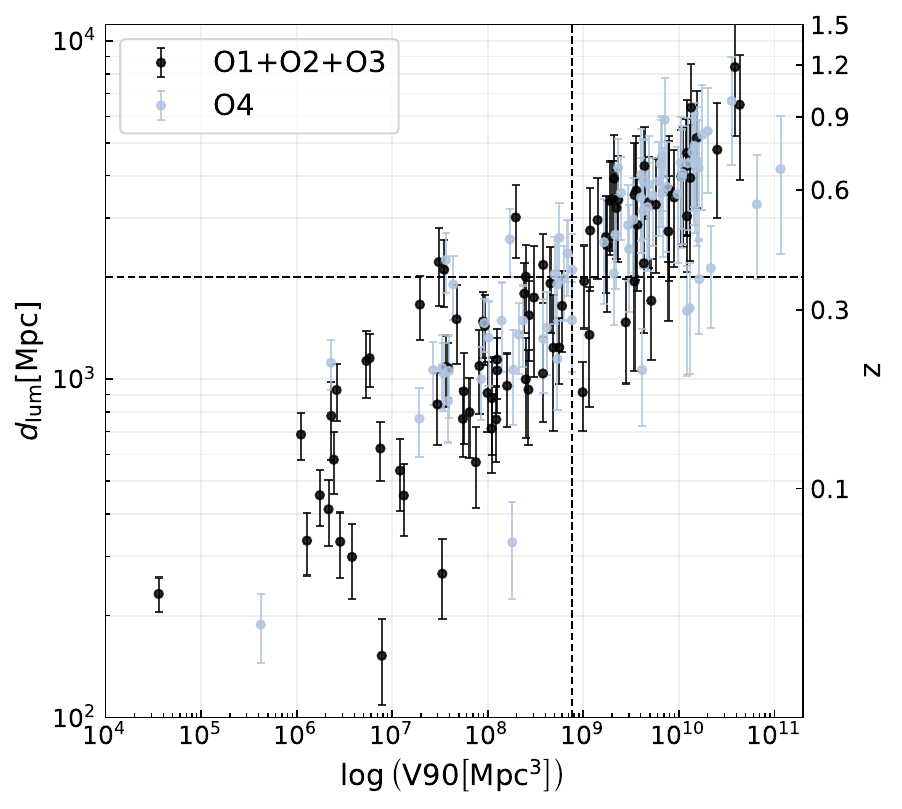}
    \caption{Luminosity distance as a function of V90 for the 159 GW events
    used in our analysis. The error bars represent an uncertainty of one standard
    deviation around the mean of the estimated distance. While the black markers
    indicate the mergers detected during the first three observing runs of the LVK
    collaboration, the light blue ones represent the events detected in O4, up to
    June 1st, 2024. The horizontal and the vertical dashed lines mark the median
    value of the luminosity distance estimates and of V90, respectively. In
    addition, we show the estimated redshift on the right axis.}
    \label{fig:dvsv90}
\end{figure}

Figure \ref{fig:dvsv90} shows the luminosity distance of all the 159 GW
events used in this work as a function of V90. The median value of the
luminosity distance (2012.3 Mpc) and of V90 (7.67$\cdot10^8{\rm Mpc}^3$)
are indicated by a horizontal dashed line and by a vertical one, respectively.


\subsection{Quaia AGN catalogue}
In order to obtain tight constraints of $f_{\rm AGN}$ with methods based
on spatial correlation like the one used in this work, is important to use
an AGN catalogue that has the highest possible value of completeness. This
property of a catalogue is in general a function of the redshifts, the
sky positions, and the luminosities of the objects it contains.

In the analysis here presented, the Quaia AGN catalogue \citep{StoreyFisher24}
\footnote{All the data concerning the Quaia catalogue, and its modeled
selection function are publicly available at
\url{https://zenodo.org/records/8060755.}} is cross-matched with the sky maps
of the GW events detailed in Section \ref{sec:GWs}. This is done once the
completeness of such a catalogue has been estimated. Quaia is an all-sky
catalogue is based on sources identified as quasar candidates in the
third data release of the Gaia mission \citep{Gaia23}. The final version of
this catalogue is obtained by selecting the objects that have an infrared
counterpart in the unWISE catalogue \citep{Lang14,Meisner19} and by performing
cuts in colours and proper motion in order to decrease the amount of
contaminants \citep[see Section 3.1 of][for details]{StoreyFisher24}. The result
is a catalogue containing 1,295,502 quasar candidates with a magnitude in the
Gaia G-band ${\rm mag}_{\rm G}<20.5$. The redshifts of all these objects are
estimated using a $k$-Nearest Neighbors model trained on the AGN present in
Quaia of which the spectra are in the 16th Data Release (DR16Q) of the the
Sloan Digital Sky Survey (SDSS) \citep{lyke20}.

\begin{figure}
    \centering
    \includegraphics[trim= 0 0 0 0 ,clip,width=1\columnwidth]{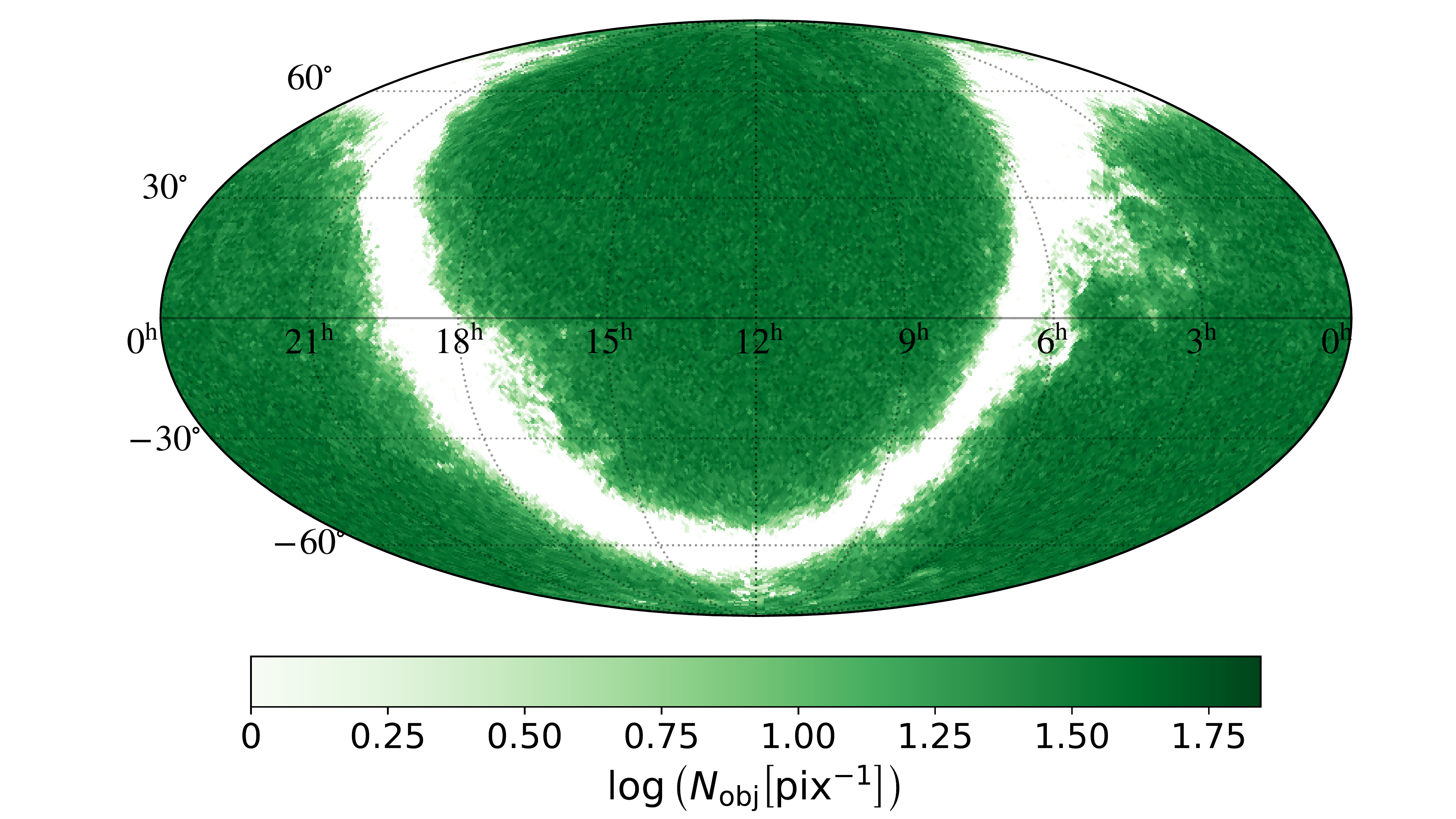}
    \caption{Mollweide projection of the sky positions in equatorial
    coordinates of the 1,295,502 objects contained in the Quaia catalogue
    with a magnitude ${\rm mag}_{\rm G}<20.5$. Different shades of green
    represent different sky-projected number densities. The resolution is
    the one of an HealPix map with NSIDE=64.}
    \label{fig:quaia}
\end{figure}

Figure \ref{fig:quaia} shows the mollweide projection of the sky distribution
of the AGN contained in the Quaia catalogue. The colour of each pixel depends
on the amount of objects in the correspondent position in the sky. The size
of each pixel is $\approx2.557\cdot10^{-4}$ steradians. Its remarkable uniformity
outside the region containing the Milky Way Galactic plane makes Quaia a
useful tool for spatial correlation analyses between AGN and GW events.

The AGN here used during the cross-match with the GW sky maps are the 660,031
objects present in Quaia that have a redshift estimate not larger than
$z=1.5$ and that are in the region of the sky where the galactic latitude
$b$ is either higher than $10^\circ$ or smaller than $-10^\circ$. The redshift
selection is performed since all the GW events used in this work are
expected to have taken place in the $0\leq z\leq1.5$ range (see Figure
\ref{fig:dvsv90}). We choose not to use the objects with a small absolute
value of galactic latitude for simplicity. We consider the density of objects
in the catalogue to be independent from their Right Ascension or Declination,
as long as they are outside of the region of the Galactic plane.

The selection function marginalised over redshift that has been modeled
for Quaia highlights density variations in regions of the sky outside the
Galactic plane \citep[See][Figure 13 ]{StoreyFisher24}. In this work thes
selection effects are not taken into account, and we obtain estimates of the
completeness of Quaia using the sky-averaged number density of objects with
the only exclusion of the Galactic plane (see Section \ref{subsec:compl}).
This choice is not expected to have a significant effect on our final results,
because during the likelihood-maximisation process we use the average value
of the catalogue's completeness inside the 90 per cent CL localisation volume
for each GW event, and the sky projection of such regions is usually much
greater than the typical angular size of the modeled number density variations.

The exclusion of the Galactic plane from Quaia removes only $\approx 1.6$
per cent of all the objects within $z=1.5$ and therefore is not expected to
decrease notably the constraining power of our analysis.

\subsection{Completeness estimation}
\label{subsec:compl}
The likelihood maximisation method presented in this work requires an estimate
of the completeness of the AGN catalogue that is used. We therefore estimate
this property of our Quaia AGN sample as a function of redshift and luminosity.
We do so in three steps:
\begin{itemize}
    \item Estimation of the bolometric luminosity of each object in Quaia with
    a redshift estimate not higher than $z=1.5$. This sub-section of Quaia
    is later referred to as ${\rm Quaia}_{{\rm z}<1.5}$, and it differs from
    our sample used during the cross-match with GW sky maps only because in the
    latter the objects in the Galactic plane region have been removed. In
    Section \ref{subsubsec:bollumcalc} we describe how these luminosities are
    evaluated starting from the magnitudes in the Gaia ${\rm G}_{\rm RP}$ band
    (${\rm mag}_{\rm G_{{\rm RP}}}$);
    \item Comparison of the bolometric luminosities obtained from the Gaia
    measurements with the ones calculated from SDSS data and presented in
    \citet{Wu22}. We do this for all the objects that are both in our sample
    of Quaia and in SDSS DR16Q, and then use the results of this comparison
    to correct the luminosity of each object in our catalogue. This step is
    done in order to have estimates of the luminosities that are compatible
    with the ones of SDSS, which is the survey used during the fit of the
    luminosity function we compare the number of objects in our catalogue with
    to calculate its completeness of our catalogue. The comparison between the
    luminosity estimates obtained from Gaia data and the ones from SDSS is
    detailed in Section \ref{subsubsec:compwu}.;
    \item Comparison between the number of objects present in our quasar sample
    in a specific redshift bin and the ones predicted by the AGN luminosity
    function presented in \citet{Kulkarni19}. This is done five times,
    considering each time a different threshold in bolometric luminosity. The
    comparison with the quasar luminosity function and the final calculation of the
    completeness of our catalogue are delineated in Section \ref{subsubsec:compkulk}.
\end{itemize}

\subsubsection{Evaluation of bolometric luminosity from ${\rm mag}_{\rm G_{{\rm RP}}}$:}
\label{subsubsec:bollumcalc}
To obtain an estimate of the bolometric luminosity of all the objects contained
in ${\rm Quaia}_{{\rm z}<1.5}$, we first calculate the flux density in
photo-electrons per second using the following standard relation:
\begin{equation}
    f_{\nu_{{\rm G}_{\rm RP}}}\left[\frac{e^-}{s}\right]=10^{-\frac{m_{{\rm G}_{\rm RP,VEG}}-{\rm ZP}_{\rm RP,VEG}}{2.5}}\ \ ,
    \label{eq:magtoflux}
\end{equation}
where ${\rm ZP}_{\rm RP,VEG}=24.7479$ is the photometric zero-point in the
Vega system for the Gaia ${\rm G}_{\rm RP}$ band. The flux density in Jansky
is then calculated by multiplying $f_\nu\left[\frac{e^-}{s}\right]$ by the
conversion factor $c_\nu=3.299\cdot10^{-36}{\rm Jy}\cdot {\rm s}/e^-$. The
values of ${\rm ZP}_{\rm RP,VEG}$ and of $c_\nu$ are taken from the online
documentation regarding the calibration of Gaia data
\footnote{\url{https://www.cosmos.esa.int/web/gaia-users/archive}}.

The intrinsic luminosity of each AGN emitted at a rest-frame frequency of
$\nu_{{\rm G}_{\rm RP}}=10^{14.588}{\rm Hz}$ \footnote{this value has been
calculated assuming a fiducial wavelength for the ${\rm G}_{\rm RP}$ band
of 7750$\AA$.} is then calculated as follows:
\begin{equation}
    \nu_{{\rm G}_{\rm RP}}L_{\nu_{{\rm G}_{\rm RP}}}=\nu_{{\rm G}_{\rm RP}}f_{\nu_{{\rm G}_{\rm RP}}}\left(4\pi d_{\rm lum}^2\right)\left(1+z\right)^{-0.657}\ \ ,
    \label{eq:fluxtolum}
\end{equation}
where $d_{\rm lum}$ is the luminosity distance correspondent to the
redshift $z$ of each object. The last term of Equation \ref{eq:fluxtolum}
is used to take into account the shape of the typical Spectral Energy Distribution
(SED) of the objects in Quaia. The value of the exponent is calculated from
a linear fit of the mean SED of all SDSS quasars presented in \citet{Richards06}.
In particular,
in the range between $\nu_{{\rm G}_{\rm RP}}$ and
$\left(1+z_{\rm max}\right)\nu_{{\rm G}_{\rm RP}}$ (where, for
${\rm Quaia}_{{\rm z}<1.5}$, $z_{\rm max}=1.5$), we find that:
\begin{equation}
    \log\left(\nu L_{\nu}\right)\propto\log\left(\nu\right)\cdot0.657\ \ ,
\end{equation}
and therefore
\begin{equation}
    \nu_{\rm em} L_{\nu_{\rm em}}= \nu_{{\rm G}_{\rm RP}}L_{\nu_{{\rm G}_{\rm RP}}}\left(1+z\right)^{0.657}\ \ ,
\end{equation}
where $\nu_{\rm em} L_{\nu_{\rm em}}$ is the luminosity emitted in the rest-frame
frequency that is observed at $\nu_{{\rm G}_{\rm RP}}$.

The bolometric luminosity is then calculated by multiplying
$\nu_{{\rm G}_{\rm RP}}L_{\nu_{{\rm G}_{\rm RP}}}$ by the value that the
frequency-dependent bolometric correction presented in \citet{Richards06} has
at the frequency $\nu_{{\rm G}_{\rm RP}}$. This value is 11.004.

\subsubsection{Comparison with the bolometric luminosity estimates
of \citet{Wu22}}
\label{subsubsec:compwu}
After computing the bolometric luminosities from the Gaia RP magnitudes
following the procedure described in Section \ref{subsubsec:bollumcalc},
we can now compare to the ones listed in the catalogue presented in
\citet{Wu22}. This catalogue contains continuum and emission-line properties
of the 750,414 broad-line Quasars of SDSS DR16Q. Among these properties
there is the bolometric luminosity of each AGN, estimated from the continuum
luminosity at the rest-frame wavelengths of 5100, 3000, and 1350 $\AA$. We
perform a cross-match between this catalogue and ${\rm Quaia}_{{\rm z}<1.5}$,
adopting to arcsec as matching radius. A total of 136,368 matches are found.
For each of these AGN it is possible to calculate the difference between the
logarithm of the bolometric luminosity estimated from ${\rm mag}_{\rm G_{{\rm RP}}}$
and the logarithm of the bolometric luminosity taken from the catalogue of
\citet{Wu22}. The mean value of the distribution of this differences is 0.073,
and its standard deviation is 0.213. The bolometric luminosities obtained from
${\rm mag}_{\rm G_{{\rm RP}}}$ are on average slightly over-estimated. We then
proceed to correct this difference.

We divide the AGN that are both in ${\rm Quaia}_{{\rm z}<1.5}$ and in the
catalogue presented in \citet{Wu22} into five bolometric luminosity bins.
This partition is performed using the estimates calculated starting from
${\rm mag}_{\rm G_{{\rm RP}}}$. Each of these sub-samples is then divided
in redshift bins, using the Quaia estimates of redshift. For each sub-sample
in luminosity we choose a different number of linear redshift bins and a
different value of the maximum redshift to consider, to ensure that in each
of these bins there are at least 10 objects. In Table \ref{tab:bins} we
list the minimum and the maximum bolometric luminosity for each of the five
sub-samples, the number of redshift bins in which it is divided, and the maximum
redshift considered in this division.
\begin{table}
    \caption{Partition of the AGN contained both in ${\rm Quaia}_{{\rm z}<1.5}$
    and in the catalogue presented in \citet{Wu22}. This partition has been used
    to compare the estimates of bolometric luminosity. We list the minimum and
    the maximum luminosity of each sub-sample. To perform this subdivision we
    use the bolometric luminosity estimates that are obtained starting from
    ${\rm mag}_{\rm G_{{\rm RP}}}$. In order to correct these estimates according to
    the values contained in the catalogue of \citet{Wu22}, we divide each sub-sample
    of ${\rm Quaia}_{{\rm z}<1.5}$ in linear redshift bins. The number of these
    bins for each sub-sample and the maximum value of the redshift used in this
    division, which has been done using the Quaia redshift estimates, are listed
    in the last two columns.}
    \begin{tabular}{c c c c}
    \hline
     $\log\left(L_{\rm bol,min}\left[{\rm erg\ s}^{-1}\right]\right)$ & $\log\left(L_{\rm bol,max}\left[{\rm erg\ s}^{-1}\right]\right)$ & ${N}_{\rm bins,z}$ & ${\rm z}_{\rm max}$\\
     \hline
        46.5 & - & 4 & 1.5 \\
        46 & 46.5 & 5 & 1.5 \\
        45.5 & 46 & 8 & 1.5 \\
        45 & 45.5 & 9 & 1.1 \\
        - & 45 & 3 & 0.5 \\
    \hline
    \end{tabular}
    \label{tab:bins}
\end{table}
    
For each bin in luminosity and redshift, we perform a linear fit between
the logarithm of the bolometric luminosities estimated starting from
${\rm mag}_{\rm G_{{\rm RP}}}$, and the one of the bolometric luminosities
taken from the catalogue of \citet{Wu22}. We therefore obtain one set of
best-fit parameters for each bin in luminosity and redshift. We finally
use these best-fit parameters to correct the bolometric luminosity estimated
from ${\rm mag}_{\rm G_{{\rm RP}}}$ of each object contained in our quasar
sample. These adjusted estimates are the ones used in the rest of the
analysis presented in this paper. The distribution of the difference between
the logarithm of the original bolometric luminosities obtained from the Gaia
magnitudes and the logarithm of their corrected version has a mean of
$0.073$ and a standard deviation of $0.027$.

\subsubsection{Comparison with the AGN luminosity function of \citet{Kulkarni19}}
\label{subsubsec:compkulk}
In order to have an estimate of the completeness of our Quaia sample as a
function of redshift we compare the number of objects it contains with the
expectation value calculated from the AGN luminosity function presented in
\citet{Kulkarni19}. In particular we use the best-fit double power law
function in which the parameters evolve as a function of redshift according to
Model 1 \citep[for the analytical expression see Equations 7, 13, 16, 17, and
18 of][while the best-fit parameters are listed in the first column of
Table 3 of the same paper]{Kulkarni19}.

We compare the observed number of AGN with the predicted one for five different
sub-samples of the catalogue. Each of these sub-samples is characterised
uniquely by a different threshold in bolometric luminosity. We divide all the
sub-samples into 8 linear redshift bins between $z=0$ and $z=1.5$. We calculate
the expected number of objects in each bin by integrating the luminosity function
of \citet{Kulkarni19}. We do this for each value of luminosity threshold.
The luminosity function is expressed as a function of the absolute monochromatic
AB magnitude at a rest frame of 1450$\AA$. For this reason, in order to obtain the
expected number of objects as a function of redshift for a specific threshold of
bolometric luminosity, we first convert this luminosity in the UV rest frame
magnitude. We do this using the magnitude-dependent bolometric correction function
presented in \citet{Runnoe12}.

\begin{figure*}
    \centering
    \includegraphics[trim= 0 0 0 0 ,clip,width=0.9\textwidth]{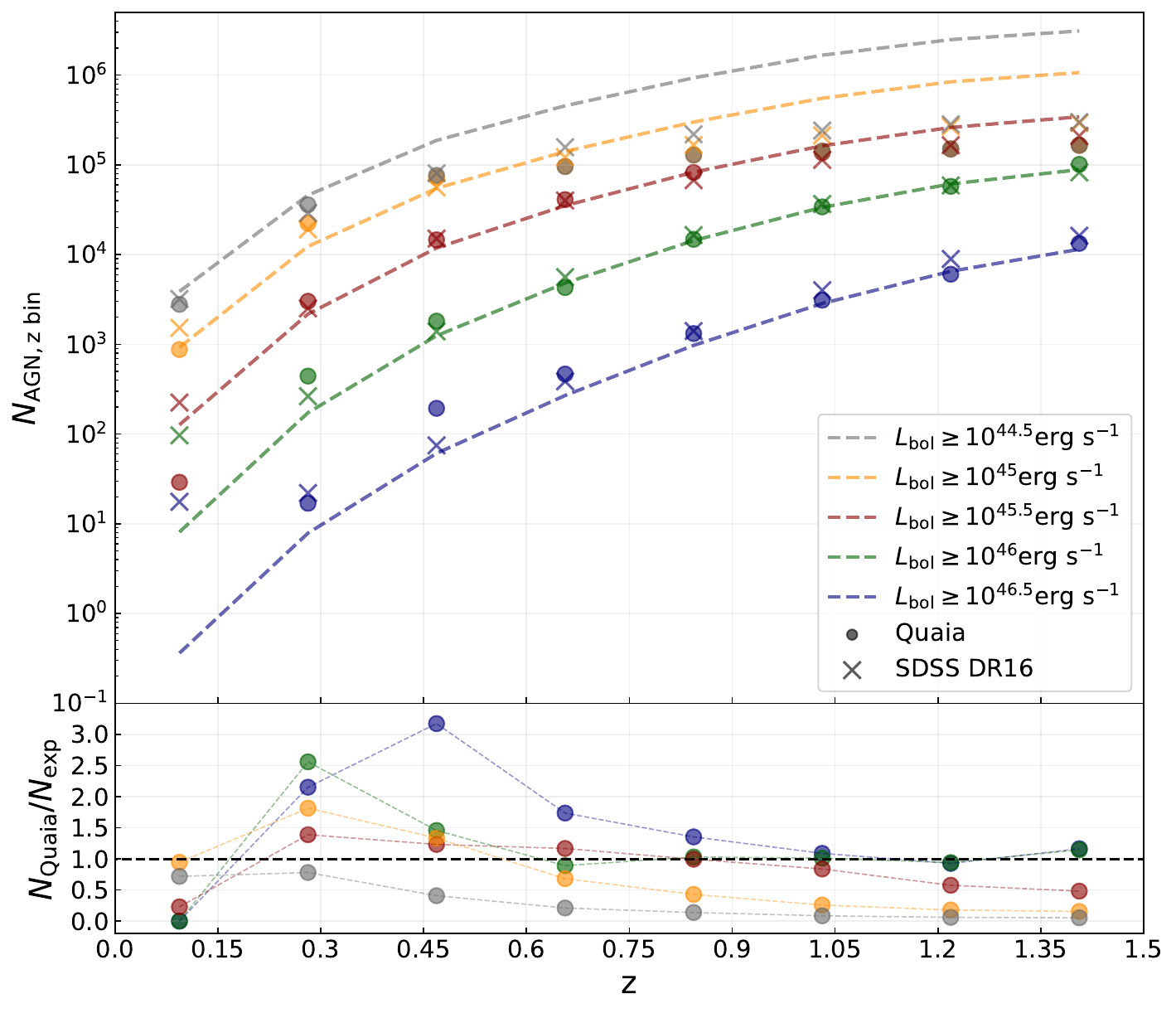}
    \caption{\textit{Top panel:} Comparison between the expected number of AGN
    in specific redshift bins as a function of different bolometric luminosity
    thresholds (dashed lines) and the observed number of objects contained in
    our Quaia sample (round markers) and in SDSS DR16 (cross markers). The expected
    numbers have been calculated integrating the redshift-evolving luminosity
    function presented in \citet{Kulkarni19}. Only the objects that have an
    absolute value of galactic latitude greater than $10^\circ$ are in our
    catalogue, and the area of the SDSS footprint is $\approx22.7$ per cent of
    the entire sky area. The numbers of observed AGN have been renormalized
    according to this. \textit{Bottom panel:} Ratio between the number of AGN
    in Quaia ($N_{\rm Quaia}$) and the expected one ($N_{\rm exp}$), as a function
    of redshift. Different colors
    represent the same different luminosity thresholds of the top panel. The
    measured decrease of the completeness as a function of redshift above
    $z\approx0.3$, and the fact that this value goes significantly below unity
    for the sub-samples with the three lowest luminosity thresholds, are due to
    the flux limit of Gaia. We consider the completeness of our catalogue to
    be 100 per cent in each bin where $N_{\rm Quaia}\geq N_{\rm exp}$.}
    \label{fig:comparisonwithkulk}
\end{figure*}

The dashed lines in Figure \ref{fig:comparisonwithkulk} show linear interpolations
of the number of expected AGN in all the different redshift bins, obtained through
the integration of the luminosity function. Different colours correspond to different
bolometric luminosity thresholds. The round markers correspond to the number of
objects brighter than the same threshold luminosities that are contained in
our quasar sample in the different redshift bins, divided by
$1-\sin\left(10^\circ\right)$ to take into account our cut in galactic latitude.
For comparison, the crosses show the number of objects in SDSS DR16Q for each
redshift bin and luminosity threshold, divided by $A_{\rm SDSS}/A_{\rm sky}$,
where $A_{\rm SDSS}=9,376\ {\rm deg}^2$ is the area of the footprint of SDSS,
and $A_{\rm sky}=41,253\ {\rm deg}^2$ is the total area of the sky. We take the
bolometric luminosities of the AGN in SDSS DR16Q from the catalogue presented
in \citet{Wu22}.

We can now compute the completeness for each redshift bin and for each
different cut in bolometric luminosity dividing the sky-area-corrected
number of observed AGN by the expected one. Whenever the former is greater
than the latter, the completeness is set to one. Because of the selection
in galactic latitude we perform, the completeness in the region where
$|b|<10^\circ$ is zero. The values of the estimated completeness for all
the different redshift bins and the different bolometric luminosity thresholds
are listed in Table \ref{tab:completeness}.
\begin{table*}
    \caption{Estimated completeness of our Quaia sample in the region of
    the sky with a galactic latitude greater than $10^\circ$ or smaller than
    $-10^\circ$. All the values are between 0 and 1, rounded up to the third
    decimal digit, and are listed for 8 linear redshift bins as a function of 5
    different bolometric luminosity thresholds.}
    \begin{adjustbox}{width=1\textwidth}
    \begin{tabular}{l c c c c c}
    \hline
    & $\log\left(L_{\rm bol}\left[{\rm erg \ s}^{-1}\right]\right)\geq46.5$ & $\log\left(L_{\rm bol}\left[{\rm erg \ s}^{-1}\right]\right)\geq46.0$ & $\log\left(L_{\rm bol}\left[{\rm erg \ s}^{-1}\right]\right)\geq45.5$ & $\log\left(L_{\rm bol}\left[{\rm erg \ s}^{-1}\right]\right)\geq45$ &
    $\log\left(L_{\rm bol}\left[{\rm erg \ s}^{-1}\right]\right)\geq44.5$\\
     \hline
        $0.0000<z\leq0.1875$ & 0 & 0 & 0.229 & 0.945 & 0.718 \\
        $0.1875<z\leq0.3750$ & 1 & 1 & 1 & 1 & 0.781 \\
        $0.3750<z\leq0.5625$ & 1 & 1 & 1 & 1 & 0.408 \\
        $0.5625<z\leq0.7500$ & 1 & 0.891 & 1 & 0.681 & 0.211 \\
        $0.7500<z\leq0.9375$ & 1 & 1 & 0.994 & 0.429 & 0.138 \\
        $0.9375<z\leq1.1250$ & 1 & 1 & 0.837 & 0.258 & 0.085 \\
        $1.1250<z\leq1.3125$ & 0.927 & 0.940 & 0.576 & 0.179 & 0.060 \\
        $1.3125<z\leq1.5000$ & 1 & 1 & 0.482 & 0.155 & 0.053 \\
    \hline
    \end{tabular}
    \end{adjustbox}
    \label{tab:completeness}
\end{table*}


\section{Method}
\label{sec:method}
Once the bolometric luminosity has been calculated for each AGN of
our catalogue and the completeness of the latter has been estimated
as a function of redshift and luminosity, we calculate the posterior
probability distribution on $f_{\rm AGN}$. To do so, we use a likelihood
function similar to the ones used in \citet{Veronesi23} and in \citet{Veronesi24}.
The general analytical expression of this function is the following:
\begin{align}
    \mathcal{L}&\left(f_{\rm AGN}\right)=\prod_{i=1}^{N_{\rm GW}}\mathcal{L}_i\left(f_{\rm AGN}\right) \nonumber \\
    &=\prod_{i=1}^{N_{\rm GW}}\left[c_i\cdot0.90\cdot f_{\rm AGN}\cdot\mathcal{S}_i
    +\left(1-c_i\cdot0.90\cdot f_{\rm AGN}\right)\mathcal{B}_i\right]\ \ ,
    \label{eq:like}
\end{align}
where $N_{\rm GW}=159$ is the total number of mergers of binaries of compact
objects we consider in the analysis, and $c_i$ is the average completeness of
the catalogue in the region occupied by the 90 per cent CL localisation volume
of the $i$-th GW event. The 0.9 factor that multiplies $f_{\rm AGN}$ is used
to take into account that we use 90 per cent CL localisation volumes, therefore
only in the 90 per cent of the time the true sources are expected to be within
them.

\begin{figure}
    \centering
    \includegraphics[trim= 0 0 0 0 ,clip,width=1.\columnwidth]{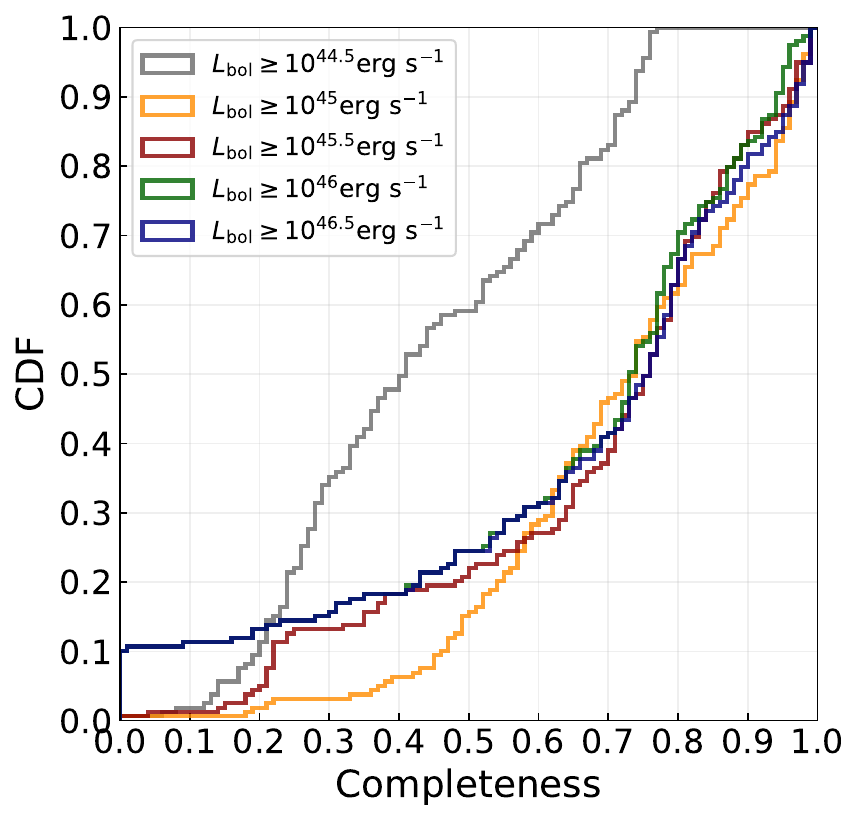}
    \caption{CDF of the values of the average completeness of our AGN catalogue
    in the region contained within the 90 per cent CL localisation volume of
    each of the 159 GW events used in this work. Different colours correspond to
    different sub-samples of the catalogue. Each of these sub-samples is uniquely
    characterised by a different value of bolometric luminosity threshold.}
    \label{fig:completenesscdf}
\end{figure}

For every GW event we calculate one value of $c_i$ for each different bolometric
luminosity threshold we use in our analysis. Figure \ref{fig:completenesscdf}
shows the Cumulative Distribution Function (CDF) of $c_i$ for all these different
sub-samples. Different colours indicate different luminosity thresholds.
The sub-samples with a threshold luminosity of $10^{46}{\rm erg\ s}^{-1}$ and
of $10^{46.5}{\rm erg\ s}^{-1}$, and that therefore contain only the most
luminous and rare AGN, have a null completeness at low redshift. For this
reason $\approx10$ per cent of the GW events have $c_i=0$ when those two
catalogues are considered.

In Equation \ref{eq:like}, $\mathcal{S}_i$ is the signal probability density
and is calculated as follows:
\begin{equation}
    \mathcal{S}_i=\frac{1}{{\rm V90}_i}\sum_{j=1}^{N_{{\rm AGN,V90}_i}}\frac{p_j}{n_{{\rm AGN},j}}
    \ \ ,
    \label{eq:Si}
\end{equation}
where $N_{{\rm AGN,V90}_i}$ is the number of AGN located within the 90 per cent CL
localisation volume of the $i$-th GW event, which has a size of ${\rm V90}_i$,
and $p_j$, measured in ${\rm Mpc}^{-3}$, is the probability density representing
how likely it is that the $i$-th merger has happened in the exact position of
the $j$-th AGN. Finally, $n_{{\rm AGN},j}$ is the number density of the AGN catalogue
in the redshift bin where the $j$-th AGN is. The value of this number density
for each AGN has been obtained by dividing the total number of objects in the
redshift bin in which it is located by the comoving volume enclosed in such a bin,
excluding the region in which $|b|<10^\circ$.

The background probability density is calculated as follows:
\begin{equation}
    \mathcal{B}_i=\frac{0.9}{{\rm V90}_i}\ \ ,
    \label{eq:Bi}
\end{equation}
where, in analogy to what has been done in \citet{Veronesi23} and in
\citet{Veronesi24}, the 0.9 factor ensures that $\mathcal{B}_i$ and
$\mathcal{S}_i$ have the same normalisation.

We cross-match the sky maps of the 159 GW events with the 5 different
sub-samples of the AGN catalogue separately, using the
{\tt postprocess.crossmatch} function of the package {\tt ligo.skymap}.
The results of these cross-matches are used to evaluate 
$\mathcal{L}\left(f_{\rm AGN}\right)$ using Equation \ref{eq:like}.
We then calculate the posterior probability distribution normalising
the likelihood function and assuming a uniform prior on $f_{\rm AGN}$
in the $[0,1]$ range.


\section{Results}
\label{sec:res}
\begin{figure*}
    \centering
    \includegraphics[trim= 0 0 0 0 ,clip,width=1\textwidth]{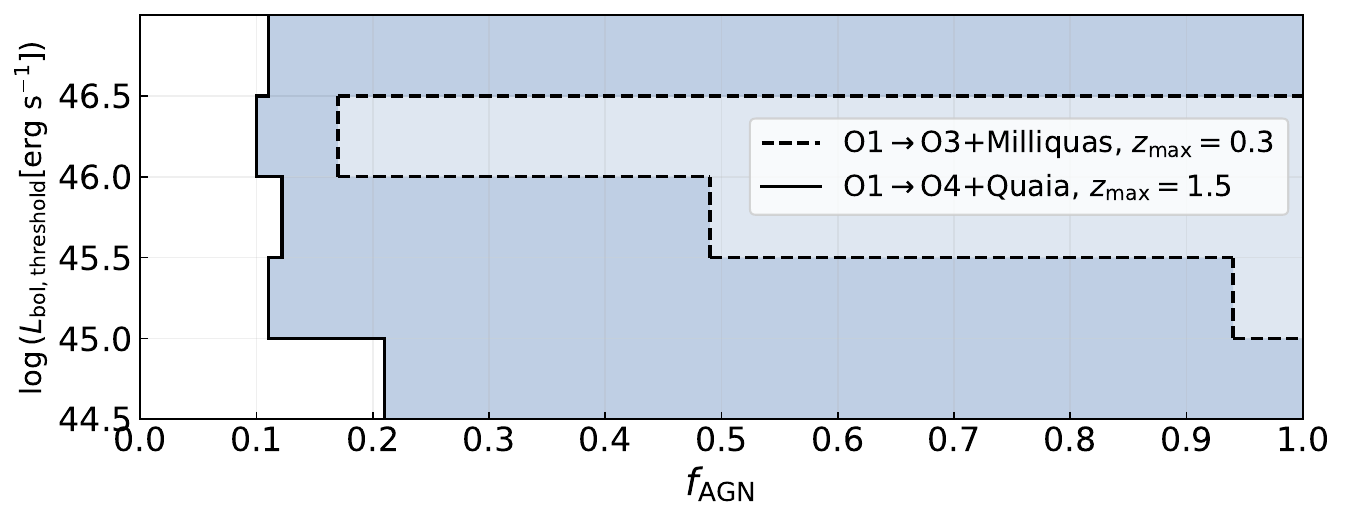}
    \caption{Observational constraints on $f_{\rm AGN}$ based on spatial
    correlation. The blue region of the plot shows the part of the parameter
    space we investigated that is rejected by our analysis at a 95 per cent
    CL. The region enclosed in in the dashed line shows the
    results obtained with a more limited dataset in \citet{Veronesi23}. Such
    previous work used the 30 GW events detected in the first three observing
    runs of the LVK collaboration that are located within $z=0.3$ at a 90 per cent
    CL, and three different catalogues of AGN in the same redshift
    range. The region enclosed in the solid line shows the results of this 
    work, which explores a wider range of AGN luminosities and uses all the
    mergers of binaries of compact objects directly detected up until June 1st, 2024.}
    \label{fig:95percentrej}
\end{figure*}

The posterior probability on $f_{\rm AGN}$ peaks at $f_{\rm AGN}=0$
independently on which bolometric luminosity threshold we consider. In
Figure \ref{fig:95percentrej} is shown in blue the region of the
investigated parameter space that our analysis rejects with a credibility
of 95 per cent.

We show the comparison with previous results, obtained in \citet{Veronesi23}
using a more limited dataset. This consisted of three catalogues, characterised
by three different cuts in bolometric luminosity, of spectroscopically
identified AGN with redshift $z\leq0.3$, that have been cross-matched with
the 30 GW events detected in the same redshift range during the first three
observing runs of the LVK collaboration. These quasar catalogues are selected
from the version 7.7b of the Milliquas catalogue \citep{flesch21}.

The increase of a factor $\approx 5$ in the number of sky maps used in the
analysis here presented and the high level of completeness of our Quaia
sample in the redshift range we consider are the causes of the increase in
the constraining power with respect to our previous work.

In Table \ref{tab:cirej} we list how many objects are considered during the
cross-matches, what fraction of our catalogue they consist of, and the
upper limits we put on $f_{\rm AGN}$ at 68, 90, and 95 per cent credibility.
\begin{table*}
    \caption{Upper limits on $f_{\rm AGN}$ we obtain at different levels of
    credibility, for the five cuts in bolometric luminosity we consider. For
    each of such cuts we also list the number of AGN used in the analysis
    ($N_{\rm AGN,cut}$), and what fraction of the total number of AGN in our
    catalogue ($N_{{\rm AGN,tot}}$) they consist of. All the values in this
    table have been rounded up to the second decimal digit.}
    \begin{adjustbox}{width=1\textwidth}
    \begin{tabular}{c c c c c c}
    \hline
     $\log\left(L_{\rm bol,threshold}\left[{\rm erg\ s}^{-1}\right]\right)$ &
     $N_{\rm AGN,cut}$ & $N_{\rm AGN,cut}/N_{{\rm AGN,tot}}$ &
     68 per cent upper limit & 90 per cent upper limit & 95 per cent upper limit\\
     \hline
        46.5 & 20,236 & 0.03 & 0.05 & 0.09 & 0.11 \\
        46 & 177,117 & 0.27 & 0.04 & 0.08 & 0.10 \\
        45.5 & 490,628 & 0.74 & 0.05 & 0.10 & 0.13 \\
        45 & 644,393 & 0.98 & 0.05 & 0.09 & 0.11 \\
        44.5 & 659,949 & 1.00 & 0.08 & 0.16 & 0.21 \\
    \hline
    \end{tabular}
    \end{adjustbox}
    \label{tab:cirej}
\end{table*}

The 95 per cent CL upper limit on the fraction of the detected GW events
that originated in an un-obscured AGN brighter than $10^{44.5}{\rm erg\ s}^{-1}$
is 21 per cent. Considering rarer, brighter quasars, the constraints become
tighter. In fact we find with the same level of credibility that no more
than 11 per cent of the mergers come from an AGN brighter than
$10^{45}{\rm erg\ s}^{-1}$. This increase of constraining power is primarily
caused by the fact that the AGN catalogue with the higher luminosity
threshold has a higher level of estimated completeness (See Figure
\ref{fig:completenesscdf}). Further increasing the value of the luminosity
threshold does not lead to significantly more stringent constraints.

\begin{figure*}
    \centering
    \includegraphics[trim= 0 0 0 0 ,clip,width=1\textwidth]{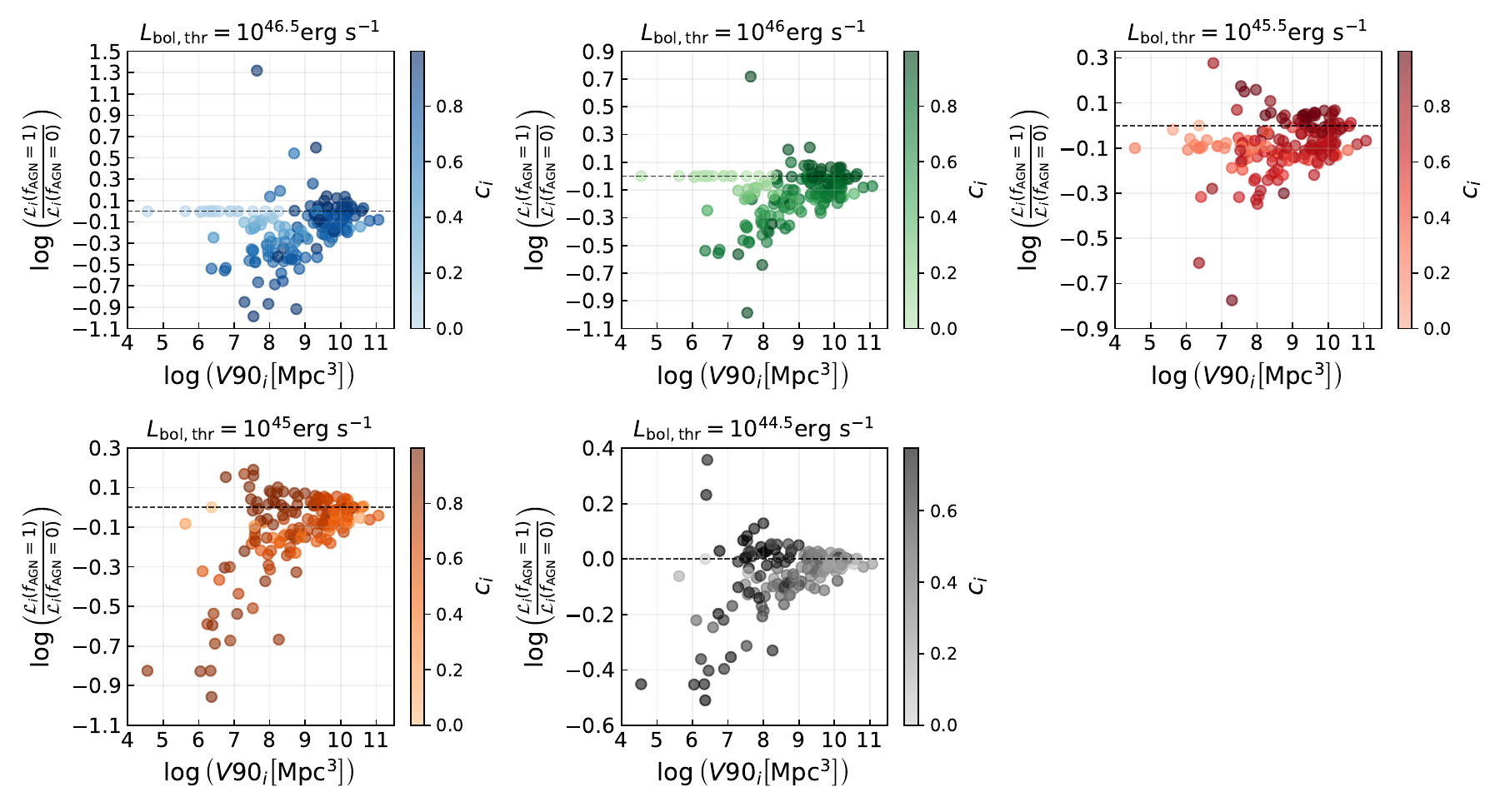}
    \caption{Logarithm of the ratio between the value of the single-event likelihood
    at $f_{\rm AGN}=1$ and the one at $f_{\rm AGN}=0$ for each detected merger, as a
    function of the corresponding V90. Each panel shows the results for a different
    bolometric luminosity threshold. In each plot, the markers are coloured as a
    function of the average completeness of the AGN catalogue in the region
    occupied by the 90 per cent CL localisation volume. The dashed
    horizontal lines mark where $\mathcal{L}_i\left(f_{\rm AGN}=1\right)=\mathcal{L}_i\left(f_{\rm AGN}=0\right)$.}
    \label{fig:difflikel}
\end{figure*}

Figure \ref{fig:difflikel} shows the logarithm of the ratio between the
single-event likelihoods ($\mathcal{L}_i$) calculated at $f_{\rm AGN}=1$ and
the ones calculated
at $f_{\rm AGN}=0$, as a function of ${\rm V90}_i$. Different panels correspond
to different bolometric luminosity thresholds. In each plot the markers are coloured
according to the average completeness inside the localisation volume of the
corresponding merger. The dashed horizontal lines indicate where the logarithm
has a null value, therefore where
$\mathcal{L}_i\left(f_{\rm AGN}=1\right)=\mathcal{L}_i\left(f_{\rm AGN}=0\right)$.
The markers above these lines correspond to GW events for which the hypothesis
of random-chance association between merger sky maps and AGN is disfavoured.

One candidate GW event in particular, S240511i, detected during the second
half of O4, corresponds to a value of
$\mathcal{L}_i\left(f_{\rm AGN}=1\right)/\mathcal{L}_i\left(f_{\rm AGN}=0\right)$
which is greater than one, independently from the luminosity threshold.
The corresponding marker is particularly evident in top part of the first
two panels of Figure \ref{fig:difflikel}. In the case of the cross-match
with the sub-sample of our catalogue with AGN brighter than
$10^{46.5}{\rm erg\ s}^{-1}$ ($10^{46}{\rm erg\ s}^{-1}$), for this GW event
$\log\left(\mathcal{L}_i\left(f_{\rm AGN}=1\right)/\mathcal{L}_i\left(f_{\rm AGN}=0\right)\right)\approx1.32\ (0.72)$.
In particular, two AGN brighter than $10^{46.5}{\rm erg\ s}^{-1}$ are found
within the 90 per cent CL localisation volume, which in the case of this GW
event has a size of ${\rm V90}\approx10^{7.64}{\rm Mpc}^3$. Rounding to the
second decimal digit, the right ascensions of the two quasars are $167.22^\circ$
and $175.58^\circ$, their declinations are $-23.64^\circ$ and $-13.35^\circ$, and
their redshift estimates from Quaia are 0.37 and 0.40. The values of
the number density of the sub-sample of the AGN catalogue with
$L_{\rm bol}\geq10^{46.5}{\rm erg\ s}^{-1}$ in the two different redshift
bins that contain the two matching quasar candidates are
$\approx10^{-8.86}{\rm Mpc}^{-3}$ and $\approx10^{-8.15}{\rm Mpc}^{-3}$.
The total probabilities within the 3D credible regions of the GW sky
map they are on the border of are $\approx0.53$ and $\approx0.77$.

The value of the single-event likelihood evaluated at $f_{\rm AGN}=1$ is
much higher than the one evaluated at $f_{\rm AGN}=0$ because two AGN are
found within the 90 per cent CL localisation volume while due to random
chance 0.06 (0.31) would be expected in the redshift bin with a number
density of $n_{\rm AGN}\approx10^{-8.86}{\rm Mpc}^{-3}$
($n_{\rm AGN}\approx10^{-8.15}{\rm Mpc}^{-3}$), considering the V90 of the
GW detection.

To estimate the statistical significance of finding a GW event with
a value of
$\mathcal{L}_i\left(f_{\rm AGN}=1\right)/\mathcal{L}_i\left(f_{\rm AGN}=0\right)$
like the ones that come from the cross-match of S240511i with
the two sub-samples of our AGN catalogue with the highest thresholds
in luminosity, we perform 500 background realisations. In each realisation
we cross-match the sky maps of the 159 detected mergers with AGN
catalogues obtained by assigning a random value of galactic longitude
to each of the objects contained in the most luminous sub-samples
of the original. We do not vary the redshifts nor the galactic latitudes
of the AGN for the sake of simplicity, since on these two spatial
coordinates depend the number density of the catalogue and its completeness,
which are parameters used during the calculation of the posterior
distribution of $f_{\rm AGN}$. In 22 (25) background realisations
involving the random AGN catalogues created from the sub-sample that
has a luminosity threshold of $10^{46.5}{\rm erg\ s}^{-1}$
($10^{46}{\rm erg\ s}^{-1}$) we find a GW event that has a single-event
likelihood ratio higher than the one S240511i has when cross-matched
with the original sub-samples. This suggests that high values of
$\mathcal{L}_i\left(f_{\rm AGN}=1\right)/\mathcal{L}_i\left(f_{\rm AGN}=0\right)$
such as the ones observed in the case of S240511i have a
probability of approximately 4.4 (5) per cent
to arise from random chance associations.

While the single-event likelihood function for S240511i favours the
hypothesis of an AGN origin, the statistical framework used in this
work is focused on analysing the entire population of GW events. The
results concerning this specific event are to be considered as hints,
not as statistically significant conclusions. Follow-up analyses,
conducted especially when the full catalogue of mergers detected during
O4 will we published, will be necessary to confidently assess whether
or not this merger has an AGN origin.

As mentioned in Section \ref{sec:intro}, the formation of binaries in
dense dynamical environment such as AGN discs  might explain the detection
of sBHs with masses the distribution of which extend to the pair-instability
mass gap region.
\begin{figure*}
    \centering
    \includegraphics[trim= 0 0 0 0 ,clip,width=1\textwidth]{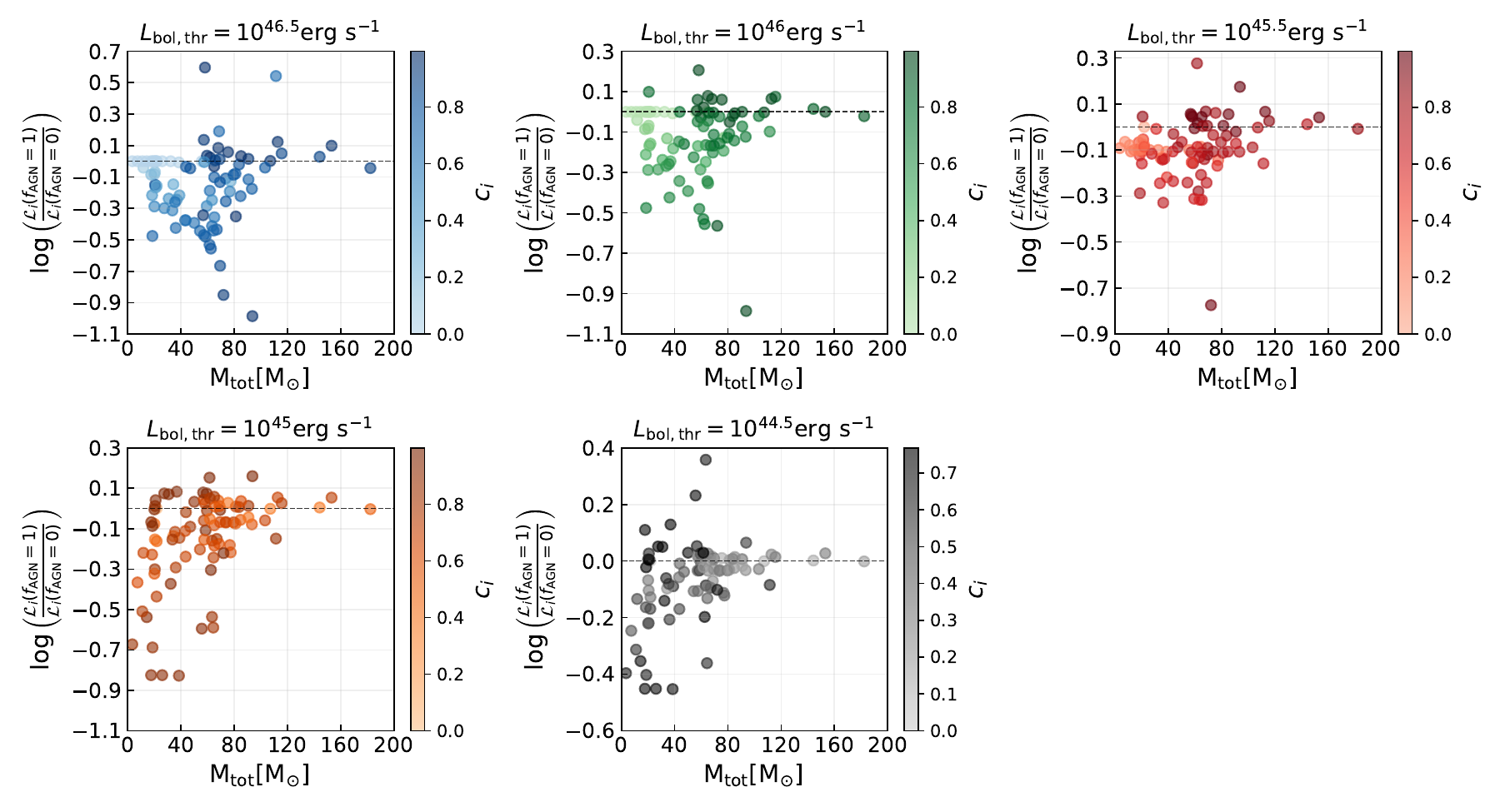}
    \caption{Logarithm of the ratio between the value of the single-event
    likelihood at $f_{\rm AGN}=1$ and the one at $f_{\rm AGN}=0$ for the
    mergers detected during the first three runs of the LVK collaboration,
    as a function of the corresponding total binary mass in the source frame.
    Each panel shows the results for a different bolometric luminosity
    threshold. In each plot, the markers are coloured as a function of the
    average completeness of the AGN catalogue in the region occupied by
    the 90 per cent CL localisation volume. The dashed horizontal lines
    mark where
    $\mathcal{L}_i\left(f_{\rm AGN}=1\right)=\mathcal{L}_i\left(f_{\rm AGN}=0\right)$.}
    \label{fig:difflikel_vsmass}
\end{figure*}
While the posterior distributions of the intrinsic binary parameters are
yet not accessible for the O4 events, for the GW events detected in the first
three observing runs it is possible to check whether there is or not a
correlation between binary mass and likelihood of being of AGN origin. The
results of this test are shown in Figure \ref{fig:difflikel_vsmass}, where
for each event the likelihood of having happened in an AGN is plotted
as a function of the total binary mass in the source frame.

Our finding is that the merging systems with the heaviest masses do not 
show a clear preference for the AGN origin. These events are the least
informative in the framework of our spatial correlation analysis. This is
because it exists a positive correlation between the total mass of the
detected merging binaries and the correspondent V90
\citep[see Figure 4 of][]{Veronesi24}, and the constraining power of
methods like the one presented in this work is mainly driven by the
best localised events \citep{Bartos17}.


\section{Discussion and conclusion}
\label{sec:concl}
In this work we present new observational constraints on the fractional
contribution of the AGN channel to the total observed merger rate of binaries
of compact objects, $f_{\rm AGN}$. These constraints are obtained using the
same spatial-correlation-based approach used in \citet{Veronesi23}. With
respect to our previous work, we make use of a new, larger dataset, which
consists of 159 GW events detected by the interferometers of the LVK
collaboration not later than June 1st, 2024, and of the all-sky AGN catalogue
Quaia. In particular we use all the AGN contained in this catalogue that have
a redshift estimate $z\leq1.5$ and an absolute value of the galactic latitude
$|b|\geq10^\circ$.

We estimate the bolometric luminosity of every AGN using the magnitudes in the
Gaia ${\rm G}_{\rm RP}$ band. We also estimate the completeness of our catalogue
as a function of redshift for different values of bolometric luminosity threshold.
The average value of the completeness within the 90 per cent CL localisation
volume of each GW event is used during the likelihood maximisation process.

We calculate the posterior probability function on $f_{\rm AGN}$ for different
thresholds on the bolometric luminosity. We find that this function always peaks
at $f_{\rm AGN}=0$. We calculate the upper limits of the 68, 90, and 95 per cent
credibility intervals on $f_{\rm AGN}$.
The main results of this work are summarised in Figure \ref{fig:95percentrej}
and in Table \ref{tab:cirej}. In particular we estimate that no more than the
21 per cent of the detected GW events used in our analysis originated from an
un-obscured AGN with a bolometric luminosity higher than $10^{44.5}{\rm erg\ s}^{-1}$
with 95 per cent credibility. Objects brighter than such threshold consist of
almost the entirety of our Quaia sample. Fainter objects are not included due to
the flux limitations of Gaia. Tighter constraints are obtained when higher luminosity
thresholds are considered. In fact we find that $f_{\rm AGN}$ is not
greater than 11 per cent at a CL of 95 per cent when un-obscured quasars brighter
than $10^{45}{\rm erg\ s}^{-1}$ are considered. The dependence of the
constraining power of our analysis with respect to the luminosity threshold
is primarily caused by the fact that more luminous AGN are rared than
dimmer ones, and spatial-correlation analyses that make use of catalogues of
rare potential host environments are characterised by lower levels of noise.

From the point of view of the theoretical modeling of binary formation
in AGN discs, there is a lack of general consensus regarding the
relation between the accretion rate or the luminosity of the host
environment and its efficiency to produce systems that merge in the
LVK band. In \citet{Yang19b}, for example, it is found that the merger
rate of sBHs increases, even if weakly, as a function of the AGN
accretion rate. On the other hand, in \citet{Grishin24} it is argued
that AGN with a luminosity $\gtrsim10^{45}{\rm erg\ s}^{-1}$ are expected
not to develop in their disc regions in which the net torque exerted
by the gas onto the compact objects is null. In these regions, usually
referred to as "migration traps", an enhanced pile-up of objects might
form in less luminous AGN, making them potentially more efficient in
forming binaries with respect to brighter hosts \citep[however, see
also][where it is argued that, due to gap-opening mechanisms induced
by the motion of the heaviest BHs in the disc, also AGN less luminous
than  $10^{43.5}{\rm erg\ s}^{-1}$ are not expected to develop
stable migration traps.]{Gilbaum24}

Thanks to the increase of a factor $\approx5$ with respect to \citet{Veronesi23}
in the number of used GW events, and to the high level of completeness of 
the quasar catalogue, we are able to put much tighter constraints on the
efficiency of the AGN channel (see Figure \ref{fig:95percentrej} for a comparison
with our previous work).

Our results have been obtained under some assumptions, which are inevitable
when inferring the properties of AGN in large catalogues. In order to calculate
the bolometric luminosities of the objects of Quaia we have to assume a shape of
the typical SED and a value for the bolometric correction. We take the values of
these AGN properties from \citet{Richards06} in order to be consistent with the
dataset presented in \citet{Wu22}, which we use to adjust the bolometric
luminosities obtained from Gaia magnitudes. Different assumptions on the SED and
on the bolometric correction might lead to different estimates of the bolometric
luminosities.

To calculate the completeness of our catalogue, we compare the observed number of
AGN with the expected one, which is obtained from the integration of a luminosity
function. We choose to use the one presented in \citet{Kulkarni19} because our
bolometric luminosities are calibrated on their SDSS-based estimates from \citet{Wu22},
and in the redshift range we consider that luminosity function has been fitted
on AGN from SDSS, which is a survey that contains un-obscured AGN detectable in the
optical band, just like the ones contained in Quaia.

An accurate estimation of the uncertainty on the completeness of the AGN catalogue
for each redshift bin and luminosity threshold is non-trivial. Factors that would
have to be taken into account are the uncertainty on the measurements of the
magnitude in the Gaia $G_{\rm RP}$ band, the potential contamination of the
catalogue with stars or galaxies, the uncertainties on the redshift
estimates, on the SED shape, and on the bolometric correction. Moreover,
the luminosity function we compare the observed number of AGN with has an intrinsic
scatter. With the only exclusion of the first redshift bin of the two sub-samples
characterised by the highest luminosity thresholds, the numbers of observed objects
used for the calculation of the completeness (see Figure \ref{fig:comparisonwithkulk})
are high enough for the systematics mentioned above to be considered the dominant
source of uncertainty, since we are in general not limited by low number counts.

As far as our analysis is concerned, taking into consideration the uncertainty
on the completeness is not expected to change the results significantly. This
is because, as described in Section \ref{sec:method}, we use the average value
of the completeness within the localisation volumes of the GW events to weight
the contribution of each of them to the total likelihood; the constraining power
of our method scales linearly with the completeness of the AGN catalogue, and
changes in this parameter are not expected to lead to any significant variation
on the value of the best estimate for $f_{\rm AGN}$. For this reason, we choose
to use the values of completeness listed in Table \ref{tab:completeness} throughout
our analysis.

Another important caveat to mention is that our analysis and the constraints we
are able to put concern un-obscured AGN, which are visible both in the optical
and in the infrared band. In order to extend our conclusions to the entire population
of AGN, one should take into account what fraction of them is not visible in the
wavelengths observed by Gaia and by WISE, which are the two surveys from which
Quaia has been created from. This obscuration fraction is in general expected to
increase as a function of redshift and to decrease as a function of luminosity \citep{Merloni14,Ueda14}.

The observational constraints presented in this work consist of a generalisation
of the ones obtained in \citet{Veronesi23}. Here we investigate a wider range of
AGN luminosities and the entirety of the redshift range reached by the
interferometers of the LVK collaboration. For this reason, in order to obtain in
the future even more general results one will need an all-sky AGN catalogue with
a lower threshold in flux with respect to Quaia, to be able to extend the analysis
to the faint-end of the un-obscured AGN population. We estimate that no more than
one GW event out of five has originated in an un-obscured AGN with a bolometric
luminosity higher than $10^{44.5}{\rm erg\ s}^{-1}$, but a larger fraction might
still come from fainter objects, or from obscured ones. This has to be taken
into consideration when comparing the results presented in this work with
constraints obtained via other techniques, like the comparison between the
observed distributions of intrinsic parameters of the merging systems and the
theoretical predictions based on physical models of binary formation in AGN discs
\citep[see e.g.][where it is estimated that the BH mass spectrum inferred from
the observed mergers listed in GWTC-2 is compatible with a fractional contribution
of the AGN channel to the total rate of 20 per cent.]{Gayathri21}

The results of this work demonstrate that spatial-correlation analyses that
use currently available data have a constraining power comparable to that of
other methods that investigate the efficiency of the AGN channel. However, in the
next years it will be possible to put even tighter constraints using the final
release of the sky maps of all the events that will be detected during the entirety
of O4 as well as the ones that will be detected during the fifth observing run of
the LVK collaboration, O5. Using more data could either reduce the upper limits on
$f_{\rm AGN}$, or it could cause a shifting of the value of such parameter that
maximised the posterior distribution, moving it away from zero.

Future developments of the statistical method used in this work involve also the
introduction of physically-motivate priors on the intrinsic binary properties. Different
binary formation channels are indeed expected to produce different features in the
distributions of the masses and the spins of the merging systems, as well as on
their eccentricity. The analysis here presented has been kept purposely agnostic as
far as the physics of the formation mechanism is concerned. Introducing
physically-motivated changes in the likelihood function might result in different
constraints on $f_{\rm AGN}$ and will inform us on which are the intrinsic binary
parameters that are able to add more information to the analysis.


\vspace{-0.8em}
\section*{Acknowledgements}
The authors thank Elia Pizzati for the stimulating discussions
regarding quasar luminosity functions and bolometric corrections,
and the anonymous referee, whose comments have helped
us to clarify the role our results play in the investigation
over the phyiscal origin of sBHs merging systems. EMR
acknowledges support from ERC Grant ``VEGA P.", number 101002511.
This research has made use of data or software obtained from
the Gravitational Wave Open Science Center (gwosc.org), a service
of LIGO Laboratory, the LIGO Scientific Collaboration, the Virgo
Collaboration, and KAGRA. LIGO Laboratory and Advanced LIGO are
funded by the United States National Science Foundation (NSF) as
well as the Science and Technology Facilities Council (STFC) of
the United Kingdom, the Max-Planck-Society (MPS), and the State
of Niedersachsen/Germany for support of the construction of Advanced
LIGO and construction and operation of the GEO600 detector.
Additional support for Advanced LIGO was provided by the Australian
Research Council. Virgo is funded, through the European Gravitational
Observatory (EGO), by the French Centre National de Recherche Scientifique
(CNRS), the Italian Istituto Nazionale di Fisica Nucleare (INFN)
and the Dutch Nikhef, with contributions by institutions from Belgium,
Germany, Greece, Hungary, Ireland, Japan, Monaco, Poland, Portugal,
Spain. KAGRA is supported by Ministry of Education, Culture, Sports,
Science and Technology (MEXT), Japan Society for the Promotion of
Science (JSPS) in Japan; National Research Foundation (NRF) and
Ministry of Science and ICT (MSIT) in Korea; Academia Sinica (AS) and
National Science and Technology Council (NSTC) in Taiwan.
{\em Software}: 
\texttt{Numpy} \citep{Harris20}; 
\texttt{Matplotlib} \citep{Hunter07}; 
\texttt{SciPy} \citep{Virtanen20};
\texttt{Astropy} \citep{Astropy13,Astropy18};
\texttt{Bilby} \citep{Ashton19};
\texttt{BAYESTAR} \citep{Singer16};
\texttt{Healpy} \citep{Zonca19}.

\vspace{-0.8em}
\section*{Data Availabilty}
The AGN catalogue used in this work, the results of the cross-matches,
and related data products are publicly available at
\url{https://zenodo.org/records/12805938}. The code used to produce the
results presented in this paper will be shared on reasonable request to
the corresponding author.

\vspace{-1.0em}
\bibliographystyle{mnras}
\bibliography{bibliography}

\begin{thebibliography}{}
\makeatletter
\relax
\def\mn@urlcharsother{\let\do\@makeother \do\$\do\&\do\#\do\^\do\_\do\%\do\~}
\def\mn@doi{\begingroup\mn@urlcharsother \@ifnextchar [ {\mn@doi@} {\mn@doi@[]}}
\def\mn@doi@[#1]#2{\def\@tempa{#1}\ifx\@tempa\@empty \href {http://dx.doi.org/#2} {doi:#2}\else \href {http://dx.doi.org/#2} {#1}\fi \endgroup}
\def\mn@eprint#1#2{\mn@eprint@#1:#2::\@nil}
\def\mn@eprint@arXiv#1{\href {http://arxiv.org/abs/#1} {{\tt arXiv:#1}}}
\def\mn@eprint@dblp#1{\href {http://dblp.uni-trier.de/rec/bibtex/#1.xml} {dblp:#1}}
\def\mn@eprint@#1:#2:#3:#4\@nil{\def\@tempa {#1}\def\@tempb {#2}\def\@tempc {#3}\ifx \@tempc \@empty \let \@tempc \@tempb \let \@tempb \@tempa \fi \ifx \@tempb \@empty \def\@tempb {arXiv}\fi \@ifundefined {mn@eprint@\@tempb}{\@tempb:\@tempc}{\expandafter \expandafter \csname mn@eprint@\@tempb\endcsname \expandafter{\@tempc}}}

\bibitem[\protect\citeauthoryear{{Abbott} et~al.,}{{Abbott} et~al.}{2017}]{Abbott17}
{Abbott} B.~P.,  et~al., 2017, \mn@doi [\prl] {10.1103/PhysRevLett.119.161101}, \href {https://ui.adsabs.harvard.edu/abs/2017PhRvL.119p1101A} {119, 161101}

\bibitem[\protect\citeauthoryear{{Abbott} et~al.,}{{Abbott} et~al.}{2023}]{Abbott23}
{Abbott} R.,  et~al., 2023, \mn@doi [Physical Review X] {10.1103/PhysRevX.13.011048}, \href {https://ui.adsabs.harvard.edu/abs/2023PhRvX..13a1048A} {13, 011048}

\bibitem[\protect\citeauthoryear{{Acernese} et~al.,}{{Acernese} et~al.}{2015}]{Acernese15}
{Acernese} F.,  et~al., 2015, \mn@doi [Classical and Quantum Gravity] {10.1088/0264-9381/32/2/024001}, \href {https://ui.adsabs.harvard.edu/abs/2015CQGra..32b4001A} {32, 024001}

\bibitem[\protect\citeauthoryear{{Akutsu} et~al.,}{{Akutsu} et~al.}{2021}]{Akutsu21}
{Akutsu} T.,  et~al., 2021, \mn@doi [Progress of Theoretical and Experimental Physics] {10.1093/ptep/ptaa125}, \href {https://ui.adsabs.harvard.edu/abs/2021PTEP.2021eA101A} {2021, 05A101}

\bibitem[\protect\citeauthoryear{Ashton et~al.}{Ashton et~al.}{2019}]{Ashton19}
Ashton G.,  et~al., 2019, \mn@doi [Astrophys. J. Suppl.] {10.3847/1538-4365/ab06fc}, 241, 27

\bibitem[\protect\citeauthoryear{{Astropy Collaboration} et~al.,}{{Astropy Collaboration} et~al.}{2013}]{Astropy13}
{Astropy Collaboration} et~al., 2013, \mn@doi [\aap] {10.1051/0004-6361/201322068}, \href {https://ui.adsabs.harvard.edu/abs/2013A&A...558A..33A} {558, A33}

\bibitem[\protect\citeauthoryear{{Astropy Collaboration} et~al.,}{{Astropy Collaboration} et~al.}{2018}]{Astropy18}
{Astropy Collaboration} et~al., 2018, \mn@doi [\aj] {10.3847/1538-3881/aabc4f}, \href {https://ui.adsabs.harvard.edu/abs/2018AJ....156..123A} {156, 123}

\bibitem[\protect\citeauthoryear{{Barrera} \& {Bartos}}{{Barrera} \& {Bartos}}{2022}]{barrera22}
{Barrera} O.,  {Bartos} I.,  2022, \mn@doi [\apjl] {10.3847/2041-8213/ac5f47}, \href {https://ui.adsabs.harvard.edu/abs/2022ApJ...929L...1B} {929, L1}

\bibitem[\protect\citeauthoryear{{Bartos}, {Haiman}, {Marka}, {Metzger}, {Stone}  \& {Marka}}{{Bartos} et~al.}{2017a}]{Bartos17}
{Bartos} I.,  {Haiman} Z.,  {Marka} Z.,  {Metzger} B.~D.,  {Stone} N.~C.,   {Marka} S.,  2017a, \mn@doi [Nature Communications] {10.1038/s41467-017-00851-7}, \href {https://ui.adsabs.harvard.edu/abs/2017NatCo...8..831B} {8, 831}

\bibitem[\protect\citeauthoryear{{Bartos}, {Kocsis}, {Haiman}  \& {M{\'a}rka}}{{Bartos} et~al.}{2017b}]{Bartos17a}
{Bartos} I.,  {Kocsis} B.,  {Haiman} Z.,   {M{\'a}rka} S.,  2017b, \mn@doi [\apj] {10.3847/1538-4357/835/2/165}, \href {https://ui.adsabs.harvard.edu/abs/2017ApJ...835..165B} {835, 165}

\bibitem[\protect\citeauthoryear{{Belczynski}}{{Belczynski}}{2020}]{Belczynski20}
{Belczynski} K.,  2020, \mn@doi [\apjl] {10.3847/2041-8213/abcbf1}, \href {https://ui.adsabs.harvard.edu/abs/2020ApJ...905L..15B} {905, L15}

\bibitem[\protect\citeauthoryear{{Belczynski} et~al.,}{{Belczynski} et~al.}{2016}]{Belczynski16}
{Belczynski} K.,  et~al., 2016, \mn@doi [\aap] {10.1051/0004-6361/201628980}, \href {https://ui.adsabs.harvard.edu/abs/2016A&A...594A..97B} {594, A97}

\bibitem[\protect\citeauthoryear{{Bellm} et~al.,}{{Bellm} et~al.}{2019}]{Bellm19}
{Bellm} E.~C.,  et~al., 2019, \mn@doi [\pasp] {10.1088/1538-3873/aaecbe}, \href {https://ui.adsabs.harvard.edu/abs/2019PASP..131a8002B} {131, 018002}

\bibitem[\protect\citeauthoryear{{Bellovary}, {Mac Low}, {McKernan}  \& {Ford}}{{Bellovary} et~al.}{2016}]{Bellovary16}
{Bellovary} J.~M.,  {Mac Low} M.-M.,  {McKernan} B.,   {Ford} K.~E.~S.,  2016, \mn@doi [\apjl] {10.3847/2041-8205/819/2/L17}, \href {https://ui.adsabs.harvard.edu/abs/2016ApJ...819L..17B} {819, L17}

\bibitem[\protect\citeauthoryear{{Cahillane} \& {Mansell}}{{Cahillane} \& {Mansell}}{2022}]{Cahillane22}
{Cahillane} C.,  {Mansell} G.,  2022, \mn@doi [Galaxies] {10.3390/galaxies10010036}, \href {https://ui.adsabs.harvard.edu/abs/2022Galax..10...36C} {10, 36}

\bibitem[\protect\citeauthoryear{{Chattopadhyay}, {Stegmann}, {Antonini}, {Barber}  \& {Romero-Shaw}}{{Chattopadhyay} et~al.}{2023}]{Chattopadhyay23}
{Chattopadhyay} D.,  {Stegmann} J.,  {Antonini} F.,  {Barber} J.,   {Romero-Shaw} I.~M.,  2023, \mn@doi [\mnras] {10.1093/mnras/stad3048}, \href {https://ui.adsabs.harvard.edu/abs/2023MNRAS.526.4908C} {526, 4908}

\bibitem[\protect\citeauthoryear{{Corley} et~al.,}{{Corley} et~al.}{2019}]{corley19}
{Corley} K.~R.,  et~al., 2019, \mn@doi [\mnras] {10.1093/mnras/stz2072}, \href {https://ui.adsabs.harvard.edu/abs/2019MNRAS.488.4459C} {488, 4459}

\bibitem[\protect\citeauthoryear{{DeLaurentiis}, {Epstein-Martin}  \& {Haiman}}{{DeLaurentiis} et~al.}{2023}]{DeLaurentiis23}
{DeLaurentiis} S.,  {Epstein-Martin} M.,   {Haiman} Z.,  2023, \mn@doi [\mnras] {10.1093/mnras/stad1412}, \href {https://ui.adsabs.harvard.edu/abs/2023MNRAS.523.1126D} {523, 1126}

\bibitem[\protect\citeauthoryear{{Fabj}, {Nasim}, {Caban}, {Ford}, {McKernan}  \& {Bellovary}}{{Fabj} et~al.}{2020}]{Fabj20}
{Fabj} G.,  {Nasim} S.~S.,  {Caban} F.,  {Ford} K.~E.~S.,  {McKernan} B.,   {Bellovary} J.~M.,  2020, \mn@doi [\mnras] {10.1093/mnras/staa3004}, \href {https://ui.adsabs.harvard.edu/abs/2020MNRAS.499.2608F} {499, 2608}

\bibitem[\protect\citeauthoryear{{Flesch}}{{Flesch}}{2021}]{flesch21}
{Flesch} E.~W.,  2021, VizieR Online Data Catalog, \href {https://ui.adsabs.harvard.edu/abs/2021yCat.7290....0F} {p. VII/290}

\bibitem[\protect\citeauthoryear{{Gaia Collaboration} et~al.,}{{Gaia Collaboration} et~al.}{2023}]{Gaia23}
{Gaia Collaboration} et~al., 2023, \mn@doi [\aap] {10.1051/0004-6361/202243940}, \href {https://ui.adsabs.harvard.edu/abs/2023A&A...674A...1G} {674, A1}

\bibitem[\protect\citeauthoryear{{Gayathri}, {Yang}, {Tagawa}, {Haiman}  \& {Bartos}}{{Gayathri} et~al.}{2021}]{Gayathri21}
{Gayathri} V.,  {Yang} Y.,  {Tagawa} H.,  {Haiman} Z.,   {Bartos} I.,  2021, \mn@doi [\apjl] {10.3847/2041-8213/ac2cc1}, \href {https://ui.adsabs.harvard.edu/abs/2021ApJ...920L..42G} {920, L42}

\bibitem[\protect\citeauthoryear{{Gerosa} \& {Berti}}{{Gerosa} \& {Berti}}{2017}]{Gerosa17}
{Gerosa} D.,  {Berti} E.,  2017, \mn@doi [\prd] {10.1103/PhysRevD.95.124046}, \href {https://ui.adsabs.harvard.edu/abs/2017PhRvD..95l4046G} {95, 124046}

\bibitem[\protect\citeauthoryear{{Gerosa} \& {Berti}}{{Gerosa} \& {Berti}}{2019}]{Gerosa19}
{Gerosa} D.,  {Berti} E.,  2019, \mn@doi [\prd] {10.1103/PhysRevD.100.041301}, \href {https://ui.adsabs.harvard.edu/abs/2019PhRvD.100d1301G} {100, 041301}

\bibitem[\protect\citeauthoryear{{Gilbaum}, {Grishin}, {Stone}  \& {Mandel}}{{Gilbaum} et~al.}{2024}]{Gilbaum24}
{Gilbaum} S.,  {Grishin} E.,  {Stone} N.~C.,   {Mandel} I.,  2024, arXiv e-prints, \href {https://ui.adsabs.harvard.edu/abs/2024arXiv241019904G} {p. arXiv:2410.19904}

\bibitem[\protect\citeauthoryear{{Graham} et~al.,}{{Graham} et~al.}{2019}]{Graham19}
{Graham} M.~J.,  et~al., 2019, \mn@doi [\pasp] {10.1088/1538-3873/ab006c}, \href {https://ui.adsabs.harvard.edu/abs/2019PASP..131g8001G} {131, 078001}

\bibitem[\protect\citeauthoryear{{Graham} et~al.,}{{Graham} et~al.}{2023}]{Graham23}
{Graham} M.~J.,  et~al., 2023, \mn@doi [\apj] {10.3847/1538-4357/aca480}, \href {https://ui.adsabs.harvard.edu/abs/2023ApJ...942...99G} {942, 99}

\bibitem[\protect\citeauthoryear{{Grishin}, {Gilbaum}  \& {Stone}}{{Grishin} et~al.}{2024}]{Grishin24}
{Grishin} E.,  {Gilbaum} S.,   {Stone} N.~C.,  2024, \mn@doi [\mnras] {10.1093/mnras/stae828}, \href {https://ui.adsabs.harvard.edu/abs/2024MNRAS.530.2114G} {530, 2114}

\bibitem[\protect\citeauthoryear{{Harris} et~al.,}{{Harris} et~al.}{2020}]{Harris20}
{Harris} C.~R.,  et~al., 2020, \mn@doi [\nat] {10.1038/s41586-020-2649-2}, \href {https://ui.adsabs.harvard.edu/abs/2020Natur.585..357H} {585, 357}

\bibitem[\protect\citeauthoryear{{Heger} \& {Woosley}}{{Heger} \& {Woosley}}{2002}]{Heger02}
{Heger} A.,  {Woosley} S.~E.,  2002, \mn@doi [\apj] {10.1086/338487}, \href {https://ui.adsabs.harvard.edu/abs/2002ApJ...567..532H} {567, 532}

\bibitem[\protect\citeauthoryear{{Hills} \& {Fullerton}}{{Hills} \& {Fullerton}}{1980}]{Hills80}
{Hills} J.~G.,  {Fullerton} L.~W.,  1980, \mn@doi [\aj] {10.1086/112798}, \href {https://ui.adsabs.harvard.edu/abs/1980AJ.....85.1281H} {85, 1281}

\bibitem[\protect\citeauthoryear{{Hopkins}, {Richards}  \& {Hernquist}}{{Hopkins} et~al.}{2007}]{Hopkins07}
{Hopkins} P.~F.,  {Richards} G.~T.,   {Hernquist} L.,  2007, \mn@doi [\apj] {10.1086/509629}, \href {https://ui.adsabs.harvard.edu/abs/2007ApJ...654..731H} {654, 731}

\bibitem[\protect\citeauthoryear{{Hunter}}{{Hunter}}{2007}]{Hunter07}
{Hunter} J.~D.,  2007, \mn@doi [Computing in Science and Engineering] {10.1109/MCSE.2007.55}, \href {https://ui.adsabs.harvard.edu/abs/2007CSE.....9...90H} {9, 90}

\bibitem[\protect\citeauthoryear{{Kulkarni}, {Worseck}  \& {Hennawi}}{{Kulkarni} et~al.}{2019}]{Kulkarni19}
{Kulkarni} G.,  {Worseck} G.,   {Hennawi} J.~F.,  2019, \mn@doi [\mnras] {10.1093/mnras/stz1493}, \href {https://ui.adsabs.harvard.edu/abs/2019MNRAS.488.1035K} {488, 1035}

\bibitem[\protect\citeauthoryear{{LIGO Scientific Collaboration} et~al.,}{{LIGO Scientific Collaboration} et~al.}{2015}]{LIGO15}
{LIGO Scientific Collaboration} et~al., 2015, \mn@doi [Classical and Quantum Gravity] {10.1088/0264-9381/32/7/074001}, \href {https://ui.adsabs.harvard.edu/abs/2015CQGra..32g4001L} {32, 074001}

\bibitem[\protect\citeauthoryear{{Lang}}{{Lang}}{2014}]{Lang14}
{Lang} D.,  2014, \mn@doi [\aj] {10.1088/0004-6256/147/5/108}, \href {https://ui.adsabs.harvard.edu/abs/2014AJ....147..108L} {147, 108}

\bibitem[\protect\citeauthoryear{{Lyke} et~al.,}{{Lyke} et~al.}{2020}]{lyke20}
{Lyke} B.~W.,  et~al., 2020, \mn@doi [\apjs] {10.3847/1538-4365/aba623}, \href {https://ui.adsabs.harvard.edu/abs/2020ApJS..250....8L} {250, 8}

\bibitem[\protect\citeauthoryear{{Mapelli}}{{Mapelli}}{2021}]{Mapelli21}
{Mapelli} M.,  2021, arXiv e-prints, \href {https://ui.adsabs.harvard.edu/abs/2021arXiv210600699M} {p. arXiv:2106.00699}

\bibitem[\protect\citeauthoryear{{Meisner}, {Lang}, {Schlafly}  \& {Schlegel}}{{Meisner} et~al.}{2019}]{Meisner19}
{Meisner} A.~M.,  {Lang} D.,  {Schlafly} E.~F.,   {Schlegel} D.~J.,  2019, \mn@doi [\pasp] {10.1088/1538-3873/ab3df4}, \href {https://ui.adsabs.harvard.edu/abs/2019PASP..131l4504M} {131, 124504}

\bibitem[\protect\citeauthoryear{{Merloni} et~al.,}{{Merloni} et~al.}{2014}]{Merloni14}
{Merloni} A.,  et~al., 2014, \mn@doi [\mnras] {10.1093/mnras/stt2149}, \href {https://ui.adsabs.harvard.edu/abs/2014MNRAS.437.3550M} {437, 3550}

\bibitem[\protect\citeauthoryear{{Nasim} et~al.,}{{Nasim} et~al.}{2023}]{Nasim23}
{Nasim} S.~S.,  et~al., 2023, \mn@doi [\mnras] {10.1093/mnras/stad1295}, \href {https://ui.adsabs.harvard.edu/abs/2023MNRAS.522.5393N} {522, 5393}

\bibitem[\protect\citeauthoryear{{Ostriker}}{{Ostriker}}{1983}]{Ostriker83}
{Ostriker} J.~P.,  1983, \mn@doi [\apj] {10.1086/161351}, \href {https://ui.adsabs.harvard.edu/abs/1983ApJ...273...99O} {273, 99}

\bibitem[\protect\citeauthoryear{{Paardekooper}, {Baruteau}, {Crida}  \& {Kley}}{{Paardekooper} et~al.}{2010}]{Paardekooper10}
{Paardekooper} S.~J.,  {Baruteau} C.,  {Crida} A.,   {Kley} W.,  2010, \mn@doi [\mnras] {10.1111/j.1365-2966.2009.15782.x}, \href {https://ui.adsabs.harvard.edu/abs/2010MNRAS.401.1950P} {401, 1950}

\bibitem[\protect\citeauthoryear{{Palmese}, {Fishbach}, {Burke}, {Annis}  \& {Liu}}{{Palmese} et~al.}{2021}]{Palmese21}
{Palmese} A.,  {Fishbach} M.,  {Burke} C.~J.,  {Annis} J.,   {Liu} X.,  2021, \mn@doi [\apjl] {10.3847/2041-8213/ac0883}, \href {https://ui.adsabs.harvard.edu/abs/2021ApJ...914L..34P} {914, L34}

\bibitem[\protect\citeauthoryear{{Planck Collaboration} et~al.,}{{Planck Collaboration} et~al.}{2016}]{Planck16}
{Planck Collaboration} et~al., 2016, \mn@doi [\aap] {10.1051/0004-6361/201525830}, \href {https://ui.adsabs.harvard.edu/abs/2016A&A...594A..13P} {594, A13}

\bibitem[\protect\citeauthoryear{{Pratten} et~al.,}{{Pratten} et~al.}{2021}]{Pratten21}
{Pratten} G.,  et~al., 2021, \mn@doi [\prd] {10.1103/PhysRevD.103.104056}, \href {https://ui.adsabs.harvard.edu/abs/2021PhRvD.103j4056P} {103, 104056}

\bibitem[\protect\citeauthoryear{{Richards} et~al.,}{{Richards} et~al.}{2006}]{Richards06}
{Richards} G.~T.,  et~al., 2006, \mn@doi [\apjs] {10.1086/506525}, \href {https://ui.adsabs.harvard.edu/abs/2006ApJS..166..470R} {166, 470}

\bibitem[\protect\citeauthoryear{{Rodriguez}, {Chatterjee}  \& {Rasio}}{{Rodriguez} et~al.}{2016}]{rodriguez16}
{Rodriguez} C.~L.,  {Chatterjee} S.,   {Rasio} F.~A.,  2016, \mn@doi [\prd] {10.1103/PhysRevD.93.084029}, \href {https://ui.adsabs.harvard.edu/abs/2016PhRvD..93h4029R} {93, 084029}

\bibitem[\protect\citeauthoryear{{Rowan}, {Boekholt}, {Kocsis}  \& {Haiman}}{{Rowan} et~al.}{2023}]{Rowan23}
{Rowan} C.,  {Boekholt} T.,  {Kocsis} B.,   {Haiman} Z.,  2023, \mn@doi [\mnras] {10.1093/mnras/stad1926}, \href {https://ui.adsabs.harvard.edu/abs/2023MNRAS.524.2770R} {524, 2770}

\bibitem[\protect\citeauthoryear{{Runnoe}, {Brotherton}  \& {Shang}}{{Runnoe} et~al.}{2012}]{Runnoe12}
{Runnoe} J.~C.,  {Brotherton} M.~S.,   {Shang} Z.,  2012, \mn@doi [\mnras] {10.1111/j.1365-2966.2012.20620.x}, \href {https://ui.adsabs.harvard.edu/abs/2012MNRAS.422..478R} {422, 478}

\bibitem[\protect\citeauthoryear{{Singer} \& {Price}}{{Singer} \& {Price}}{2016}]{Singer16}
{Singer} L.~P.,  {Price} L.~R.,  2016, \mn@doi [Physical Reviews D] {10.1103/PhysRevD.93.024013}, \href {https://ui.adsabs.harvard.edu/abs/2016PhRvD..93b4013S} {93, 024013}

\bibitem[\protect\citeauthoryear{{Storey-Fisher}, {Hogg}, {Rix}, {Eilers}, {Fabbian}, {Blanton}  \& {Alonso}}{{Storey-Fisher} et~al.}{2024}]{StoreyFisher24}
{Storey-Fisher} K.,  {Hogg} D.~W.,  {Rix} H.-W.,  {Eilers} A.-C.,  {Fabbian} G.,  {Blanton} M.~R.,   {Alonso} D.,  2024, \mn@doi [\apj] {10.3847/1538-4357/ad1328}, \href {https://ui.adsabs.harvard.edu/abs/2024ApJ...964...69S} {964, 69}

\bibitem[\protect\citeauthoryear{{Tagawa}, {Haiman}  \& {Kocsis}}{{Tagawa} et~al.}{2020}]{Tagawa20}
{Tagawa} H.,  {Haiman} Z.,   {Kocsis} B.,  2020, \mn@doi [\apj] {10.3847/1538-4357/ab9b8c}, \href {https://ui.adsabs.harvard.edu/abs/2020ApJ...898...25T} {898, 25}

\bibitem[\protect\citeauthoryear{{Ueda}, {Akiyama}, {Hasinger}, {Miyaji}  \& {Watson}}{{Ueda} et~al.}{2014}]{Ueda14}
{Ueda} Y.,  {Akiyama} M.,  {Hasinger} G.,  {Miyaji} T.,   {Watson} M.~G.,  2014, \mn@doi [\apj] {10.1088/0004-637X/786/2/104}, \href {https://ui.adsabs.harvard.edu/abs/2014ApJ...786..104U} {786, 104}

\bibitem[\protect\citeauthoryear{{Veronesi}, {Rossi}, {van Velzen}  \& {Buscicchio}}{{Veronesi} et~al.}{2022}]{veronesi22}
{Veronesi} N.,  {Rossi} E.~M.,  {van Velzen} S.,   {Buscicchio} R.,  2022, \mn@doi [\mnras] {10.1093/mnras/stac1346}, \href {https://ui.adsabs.harvard.edu/abs/2022MNRAS.514.2092V} {514, 2092}

\bibitem[\protect\citeauthoryear{{Veronesi}, {Rossi}  \& {van Velzen}}{{Veronesi} et~al.}{2023}]{Veronesi23}
{Veronesi} N.,  {Rossi} E.~M.,   {van Velzen} S.,  2023, \mn@doi [\mnras] {10.1093/mnras/stad3157}, \href {https://ui.adsabs.harvard.edu/abs/2023MNRAS.526.6031V} {526, 6031}

\bibitem[\protect\citeauthoryear{{Veronesi}, {van Velzen}  \& {Rossi}}{{Veronesi} et~al.}{2024}]{Veronesi24}
{Veronesi} N.,  {van Velzen} S.,   {Rossi} E.~M.,  2024, \mn@doi [arXiv e-prints] {10.48550/arXiv.2405.05318}, \href {https://ui.adsabs.harvard.edu/abs/2024arXiv240505318V} {p. arXiv:2405.05318}

\bibitem[\protect\citeauthoryear{{Virtanen} et~al.,}{{Virtanen} et~al.}{2020}]{Virtanen20}
{Virtanen} P.,  et~al., 2020, \mn@doi [Nature Methods] {10.1038/s41592-019-0686-2}, \href {https://ui.adsabs.harvard.edu/abs/2020NatMe..17..261V} {17, 261}

\bibitem[\protect\citeauthoryear{{Wu} \& {Shen}}{{Wu} \& {Shen}}{2022}]{Wu22}
{Wu} Q.,  {Shen} Y.,  2022, \mn@doi [\apjs] {10.3847/1538-4365/ac9ead}, \href {https://ui.adsabs.harvard.edu/abs/2022ApJS..263...42W} {263, 42}

\bibitem[\protect\citeauthoryear{{Yang} et~al.,}{{Yang} et~al.}{2019a}]{yang19}
{Yang} Y.,  et~al., 2019a, \mn@doi [\prl] {10.1103/PhysRevLett.123.181101}, \href {https://ui.adsabs.harvard.edu/abs/2019PhRvL.123r1101Y} {123, 181101}

\bibitem[\protect\citeauthoryear{{Yang}, {Bartos}, {Haiman}, {Kocsis}, {M{\'a}rka}, {Stone}  \& {M{\'a}rka}}{{Yang} et~al.}{2019b}]{Yang19b}
{Yang} Y.,  {Bartos} I.,  {Haiman} Z.,  {Kocsis} B.,  {M{\'a}rka} Z.,  {Stone} N.~C.,   {M{\'a}rka} S.,  2019b, \mn@doi [\apj] {10.3847/1538-4357/ab16e3}, \href {https://ui.adsabs.harvard.edu/abs/2019ApJ...876..122Y} {876, 122}

\bibitem[\protect\citeauthoryear{{Ziosi}, {Mapelli}, {Branchesi}  \& {Tormen}}{{Ziosi} et~al.}{2014}]{Ziosi14}
{Ziosi} B.~M.,  {Mapelli} M.,  {Branchesi} M.,   {Tormen} G.,  2014, \mn@doi [\mnras] {10.1093/mnras/stu824}, \href {https://ui.adsabs.harvard.edu/abs/2014MNRAS.441.3703Z} {441, 3703}

\bibitem[\protect\citeauthoryear{Zonca, Singer, Lenz, Reinecke, Rosset, Hivon  \& Gorski}{Zonca et~al.}{2019}]{Zonca19}
Zonca A.,  Singer L.,  Lenz D.,  Reinecke M.,  Rosset C.,  Hivon E.,   Gorski K.,  2019, \mn@doi [Journal of Open Source Software] {10.21105/joss.01298}, 4, 1298

\makeatother
\end{thebibliography}

\bsp
\label{lastpage}
\end{document}